\documentclass[titlepage,12pt,reqno]{article}
\usepackage{amsmath,amssymb,amsfonts,graphics}
\usepackage[linktocpage=true,colorlinks,pdftex]{hyperref}
\usepackage{xcolor}
\colorlet{linkequation}{blue}
\usepackage{bm}
\usepackage{doi}
\usepackage{hyperref}
\hypersetup{colorlinks, citecolor=violet, filecolor=black, linkcolor=black, urlcolor=blue}
\newcommand*{\refeq}[1]{%
  \begingroup
    \hypersetup{
      linkcolor=linkequation,
      linkbordercolor=linkequation,
    }%
    \ref{#1}%
  \endgroup
}
\allowdisplaybreaks
\newcommand{\normord}[1]{:\mathrel{#1}:}

\begin{document} 


\begin{titlepage}

\centerline{\LARGE \bf Entanglement Entropy in the $\sigma$-Model} 
\centerline{\LARGE \bf with the de Sitter Target Space}
\vskip 1.5cm
\centerline{ \bf Ion V. Vancea }
\vskip 0.5cm
\centerline{\sl Grupo de F{\'{\i}}sica Te\'{o}rica e Matem\'{a}tica F\'{\i}sica, Departamento de F\'{\i}sica}
\centerline{\sl Universidade Federal Rural do Rio de Janeiro}
\centerline{\sl Cx. Postal 23851, BR 465 Km 7, 23890-000 Serop\'{e}dica - RJ,
Brazil}
\centerline{
\texttt{\small ionvancea@ufrrj.br} 
}

\vspace{0.5cm}

\centerline{17 September 2017}

\vskip 1.4cm
\centerline{\large\bf Abstract} 
We derive the formula of the entanglement entropy between the left and right oscillating modes of the $\sigma$-model with the de Sitter target space. To this end, we study the theory in the \emph{cosmological gauge} in which the non-vanishing components of the metric on the two-dimensional base space are functions of the expansion parameter of the de Sitter space. The model is embedded in the causal north pole diamond of the Penrose diagram. We argue that the cosmological gauge is natural to the $\sigma$-model as it is compatible with the canonical quantization relations. In this gauge, we obtain a new general solution to the equations of motion in terms of time-independent oscillating modes. The constraint structure is adequate for quantization in the Gupta-Bleuler formalism. We construct the space of states as a one-parameter family of Hilbert spaces and give the Bargmann-Fock and Jordan-Schwinger representations of it. Also, we give a simple description of the physical subspace as an infinite product of $\mathcal{D}^{+}_{\frac{1}{2}}$ in the positive discreet series irreducible representations of the $SU(1,1)$ group. We construct the map generated by the Hamiltonian between states at two different values of time and show how it produces the entanglement of left and right excitations. Next, we derive the formula of the entanglement entropy of the reduced density matrix for the ground state acted upon by the Hamiltonian map. Finally, we determine the asymptotic form of the entanglement entropy of a single mode bi-oscillator in the limit of large values of time.

\vskip 0.7cm 

{\bf Keywords:} sigma model; string theory; quantum field theory in the de Sitter space; entanglement entropy.

\noindent 

\end{titlepage}


\section{Introduction}

Entanglement is a fundamental property that describes non-local correlations in a quantum system. On manifolds with horizons, the entanglement is universal and affects the ground states of the field theories defined on them \cite{Muller:1995mz}. In particular, this is also the case of the de Sitter space that presents event horizons. There, the entanglement has recently been investigate in  various quantum field theories \cite{Kabat:2002hj,Maldacena:2012xp,Nomura:2013lia,Kanno:2014lma,Iizuka:2014rua,Pavao:2016rhe}. 

In this paper, we are going to study the entanglement in the two-dimensional non-linear $\sigma$-model defined by the embedding $x:\Sigma^{1,1} \longrightarrow dS_4$, where $\Sigma^{1,1}$ is the base space of the model with the topology of a two-dimensional cylinder and $dS_4$ is the four-dimensional de Sitter target space. One reason to study the entanglement in this theory is that it has a new aspect absent in field theories defined on the de Sitter space, namely, there are left and right degrees of freedom on the base space that are entangled among themselves as a consequence of the interaction with the target space metric. A second reason for which this problem deserves attention is that it could help us formulate the (non-critical) string theory compactified to a simple space-time with a positive cosmological constant as appears to be the case of our universe according to recent astronomical observations \cite{Schmidt:1998ys,Riess:1998cb,Perlmutter:1998np}. A third reason is to improve the general knowledge of the quantum dynamics of strings in the de Sitter space that could help us understand the microscopic structure of more general time-dependent backgrounds in the string theory \cite{Maloney:2002rr,Burgess:2003ic,Kachru:2003aw} and phenomena such as the string production in the early universe \cite{Gubser:2003vk}.

A particularly useful quantity to understand the entanglement in a quantum field theory is the entanglement entropy of its ground state. Since no covariant quantization method is available to date, the most convenient way to formulate and calculate the entanglement entropy in the $\sigma$-model is by embedding the base space into a local causal region of the de Sitter space and by fixing the gauge symmetries\footnote{However, a non-linear $\sigma$-model that is ghost-free was recently constructed in \cite{deRham:2015ijs} for non-compact target spaces in the context of the massive gravity (see also \cite{deRham:2016plk}).}. Then the quantum operators on the base space can be interpreted in terms of local quantities of the de Sitter space. This idea was used previously to obtain the semi-classical mass of states, the maximum number of a single excitation and the equation of state of normal excitations around the geodesics described by the centre of mass of different string configurations \cite{deVega:1987veo,deVega:1994yz,RamonMedrano:1999gm,Bouchareb:2005ck}. The same quantities were calculated in a general time-dependent gauge of the two-dimensional Weyl symmetry for the small strings in \cite{Li:2007gf}. The one-loop divergences in the quantum Nambu-Goto model were obtained in \cite{Viswanathan:1996yg} in the path integral formalism. Also, an extensive analysis of the back-reaction on the classical fields in different types of time-dependent backgrounds and string conditions, including the de Sitter space, was done in \cite{Bozhilov:2001kw}\footnote{The string interpretation is usually made by imposing some string properties on the $\sigma$-model while ignoring the fact that the de Sitter space is not a string background as it does not satisfy the $\beta$-function equations of strings \cite{Callan:1985ia}.}.

In order to derive the entanglement entropy, we will analyse the classical and quantum dynamics of the $\sigma$-model in a new diagonal gauge in which the non-vanishing component of the two-dimensional metric is proportional to the expansion factor of the de Sitter space that will be refereed to as the cosmological gauge. We will argue that this is a natural gauge of the theory in the sense that it is the only gauge that depends just on the time coordinate and is compatible with the canonical commutation relations. By embedding the two-dimensional base space into the north pole diamond of the Penrose diagram of the de Sitter space and by choosing the co-moving frame in the Friedman-Walker coordinates, the analysis can be considerably simplified. In this embedding, the quantum theory on the two-dimensional base space and the entanglement entropy of the ground state are both local in the de Sitter space, in the sense that all quantities are defined in a neighbourhood of the embedded cylinder. However, from the point of view of the two-dimensional field theory, the entanglement entropy is a non-local quantity as usual. Compared with the previous studies, this approach to the quantization is similar to the one taken in \cite{Li:2007gf} in an arbitrary time-dependent gauge. Nevertheless, by imposing the quantization conditions, the arbitrariness of the classical time-dependent gauge parameter is restricted to the cosmological gauge parameter and that leads to a different quantum theory from the one considered there. Also, while in the previous works the system was subjected to Bogoliubov transformations meant to remove the non-diagonal terms from the Hamiltonian, in the present study we will focus on these very terms as they are responsible for the entanglement of the left and right moving modes.

It is important to note that the entropy of the entanglement of the left and right moving modes (LREE) has been studied previously in the context of boundary states in bosonic conformal field theories. In  
\cite{PandoZayas:2014wsa}, it was discovered that the system obtained by tracing out one type of moving modes of a free bosonic conformal field theory has properties similar to a conformal field gas for Dirichlet as well as Neumann boundary conditions. In 
\cite{Zayas:2016drv}, the LREE was found for WZW D-branes and in \cite{Schnitzer:2015gpa} the R\'{e}nyi entropy was used to obtain the LREE of a general Dp-brane. These results focus on the microscopic boundary states rather than general string states since in the string space-times, i. e. in which the space-time metric satisfies the corresponding $\beta$-function equation, the bosonic string fields are untangled. On the other hand, the sigma model degrees of freedom are entangled in the de Sitter space-time due to the interaction with the metric components. Therefore, we can define the LREE of the $\sigma$-model. 
Universal properties of the LREE in the two-dimensional conformal field theories have been the object of study  in \cite{Das:2015oha}. Also, components of the metric and general geometric properties have been shown to emerge from the Fisher entropy in \cite{Molina-Vilaplana:2015mja,Molina-Vilaplana:2015rra} in the context of the continuous multiscale  entanglement renormalization ansatz (cMERA) \cite{Vidal:2008zz,Haegeman:2011uy} which is related to the de Sitter space and in the Euclidean formulation that could be extended to time-dependent backgrounds \cite{Caputa:2017urj}.

The paper is organized as follows. The classical dynamics of the $\sigma$-model in the cosmological gauge is discussed in Section 2. Here, it is obtained a new general solution to the equations of motion that is a linear superposition of the Hankel functions of both types with constant coefficients which reflect the symmetry of the base space. This solution has properties similar to the one obtained for closed strings moving in null cosmology backgrounds from \cite{Madhu:2009jh} and is simpler compared with the one obtained in the case of an arbitrary time-dependent gauge in \cite{Li:2007gf}. In the same section, we discuss the independent constraints on the classical fields and argue that they can be interpreted as the canonical momentum and Hamiltonian on the base space. However, the time-translation generated by the Hamiltonian does not conserve the energy due to the interaction with the background metric. In Section 3, we quantize the $\sigma$-model by using the Gupta-Bleuler method. We will argue that the time-dependent Hilbert space of the system is a time-independent Hilbert space tensored with the smooth time-dependent functions on the north pole diamond. The physical states form a subspace of it defined as the intersection of the kernels of the momentum and Hamiltonian operators, respectively. We show that the momentum constraint acts as a level matching conditions on the time-independent Hilbert space while the Hamiltonian constraint produces a recurrence relation among the time-dependent functions. The canonical Hamiltonian presents itself as a linear combination of an infinite number of generators of the $su(1,1)$-algebra with time-dependent coefficients. These generators correspond to pairs of identical left and right modes along the same direction of the de Sitter space in the Jordan-Schwinger bi-oscillator representation. We construct the Jordan-Schwinger representation of the time-independent Hilbert space and determine the map between the Bargmann-Fock and the Jordan-Schwinger representations, respectively. By using this map, we calculate the explicit form of  the time-dependent Hilbert space in the positive discrete series irreducible representations $\mathcal{D}^{+}_{k^{i}_{m}}$ of the $SU(1,1)$ group \cite{Bargmann:1946me}. The physical subspace of it is 
the infinite product of $\mathcal{D}^{+}_{\frac{1}{2}}$ over all modes. This space, when tensored with the constrained time-dependent functions, provides an explicit representation of the physical states. In general, the states from the time-dependent total Hilbert space are entangled time-independent oscillator states from the left and right sectors with time-dependent coefficients. The Hamiltonian just constraints further these coefficients modifying the degree of  entanglement. Therefore, the time-dependent states are quantum quenches of the Hamiltonian. In Section 4, we will show that the normal ordered Hamiltonian generates an unitary map between the states at different times. This map is not the energy conserving evolution map of the total system and its main effect on states is to change the entanglement between the left and right oscillating modes of the state it acts on. A good measure of the degree of entanglement produced by the Hamiltonian map between two instants of time is the (time-dependent) von Neumman LREE  of the ground state. We will derive the general formula of it and determine explicitly its asymptotic form at large values of time for a single mode. The last section is devoted to discussions.   

\section{Classical Dynamics in the Cosmological Gauge}

In this section, we will derive the classical dynamics of the two-dimensional $\sigma$-model with the de Sitter target space in the cosmological gauge. The main result obtained here is the new general solution to the equations of motion presented in the equation (\refeq{solutions-x}). A similar analysis of the small string in an arbitrary time-dependent gauge that generated a different solution from ours was performed in \cite{Li:2007gf}.

Our starting point is the embedding $x: \Sigma^{1,1} \rightarrow dS_4$ of the two-dimensional base space $\Sigma^{1,1} = S^{1} \times I$, $I \subset \mathbb{R}$,  into the north pole diamond of the Penrose diagram of the four-dimensional target space $dS_4$. The dynamics of the components $x^{\mu}(\sigma)$ of the embedding is described by the action functional
\begin{equation}
S_0 [x] = - \frac{1}{2 \pi} \int d^2 \sigma \sqrt{-h(\sigma)} h^{ab}(\sigma)
\partial_a x^{\mu}(\sigma) \partial_b x^{\nu} (\sigma) g_{\mu \nu} (x),
\label{action-general}
\end{equation}
where $a, b = 0, 1$ are the base space indices that label the time-like and space-like coordinates $\sigma^0 \in I \subseteq \mathbb{R}^{+}$ and $\sigma^1 \in [0,2 \pi]$, respectively, $h_{ab}(\sigma) = h_{ab}(\sigma^0,\sigma^1 )$ is the metric on $\Sigma^{1,1}$ of signature $(-,+)$ and $h(\sigma) = \det h_{ab}(\sigma) $. The time-dependent metric on $dS_4$ is $g_{\mu \nu}(t)$ and $\mu , \nu = 0, \ldots , 3$ are the target space indices. In the north pole diamond of the de Sitter space diagram, one can conveniently chose the planar coordinates in the co-moving frame in which the local line element takes the following form
\begin{equation}
ds^2 = - dt^2 + e^{2Ht}\delta_{ij} dx^i dx^j .
\label{de-Sitter-metric}
\end{equation}
Here, $H$ is the Hubble constant and $i,j = 1, 2, 3$ label the spatial directions in the transverse sections of $dS_4$\footnote{In string theory, this setting would correspond to the string compactified on the internal $K^6$ cycles of the ten dimensional space-time with the metric
\begin{equation}
ds^2 = - dt^2 + e^{2Ht}\delta_{ij} dx^i dx^j + d^{2}_{K^6}.
\nonumber
\label{full-metric}
\end{equation}
However, no such compactification is known at present with or without fluxes or higher dimensional objects. Therefore, when the results obtained for the $\sigma$-model are interpreted in terms of strings, the compact space term from the above equation should be ignored.}. The purpose of this setting is to restrict the dynamics to a local region of the de Sitter space where there are local time-like Killing vectors with the past-to-future orientation.

The action $S_0 [x]$ is invariant under the kinematic de Sitter group in the target space and under the gauge symmetries generated by the Weyl and the reparametrization transformations of the base space. These are defined by the following relations
\begin{align}
h_{ab}(\sigma) & \rightarrow  h'_{ab}(\sigma) = e^{\lambda(\sigma)}h_{ab}(\sigma),
\label{Weyl-class}
\\
\sigma^a & \rightarrow   \sigma'^{a} = f^{a}(\sigma),
\label{reparam-class}
\end{align}
where $\lambda(\sigma)$ and $f^a (\sigma)$ are arbitrary functions on the two-dimensional coordinates, respectively. In order to study the dynamics of the fields $x^i(\sigma^{0},\sigma^{1})$, it is convenient to choose the orientation of the embedding $x$ similar to the static gauge in the Minkowski space-time by identifying the time-like coordinates on the base and target spaces, respectively,
\begin{equation}
\sigma^0 = t.
\label{static-gauge}
\end{equation}
Next, we use the gauge symmetries given by the equations (\refeq{Weyl-class}) and (\refeq{reparam-class}), respectively, to fix the components of the two-dimensional metric to the diagonal form
\begin{equation}
h(\sigma ) = -1, 
\hspace{0.5cm}
h_{01} = 0,
\hspace{0,5cm}
h_{11} = \omega (\sigma^0 = t),
\label{metric-gauge-ws}
\end{equation}
where $\omega (t)$ is an arbitrary smooth function of time only. The classical general time-dependent gauge was considered in \cite{Li:2007gf} where the dynamics of the little string was discussed thoroughly. This gauge can be fixed further to the cosmological gauge defined by the following equation
\begin{equation}
\omega (t) = \omega_0 e^{-2Ht},
\label{metric-cosmological-gauge}
\end{equation} 
where $\omega_0 $ is a constant corresponding to the initial value of time
$\sigma_0 = t_0$ that can be taken one (see the equation (\refeq{time-dep-ETC})). The gauge fixing defined by the above equation is completely arbitrary in the classical theory. However, a simple argument presented in the Appendix A shows that the classical gauge functions $\omega (t)$ are naturally restricted to the cosmological gauge parameter by the canonical quantization procedure. Therefore, in the rest of this paper we will work in the cosmological gauge.
  
The gauge fixed action can be obtained by plugging the equations (\refeq{metric-gauge-ws}) and (\refeq{metric-cosmological-gauge}) into the equation (\refeq{action-general}) with the following result
\begin{equation}
S[x] =  \frac{1}{2 \pi} \int dt d\sigma
\left[ 
- e^{-2Ht} + \left( \partial_t x^{i} (t,\sigma) \right)^2
- e^{2Ht} \left( \partial_{\sigma} x^{i} (t,\sigma) \right)^2
\right],
\label{action-cosmo-gauge}
\end{equation}
where $\sigma$ stands now for $\sigma^{1}$. The variation of $S[x]$ with respect to the fields $x^i(t,\sigma)$ produces the following equations of motion
\begin{equation}
\partial^{2}_{t} x^i (t,\sigma) - 
 e^{4Ht}\partial^{2}_{\sigma}
x^i (t,\sigma) = 0.
\label{eq-motion-x}
\end{equation} 
On the other hand, by varying $S[x]$ with respect to the metric components, three equations are generated of which only two are independent of each other, namely
\begin{align}
\mathit{P} (t, \sigma) & =  \frac{1}{8 \pi}\delta_{ij}
\partial_{t} x^i (t,\sigma) \partial_{\sigma} x^j (t,\sigma) 
\simeq 0,
\label{constr-level-matching}
\\
\mathit{H} (t, \sigma) & =  \frac{1}{8 \pi }\left[- e^{-4Ht} + e^{-2Ht} \left( \partial_t x^{i} (t,\sigma) \right)^2
+ e^{2Ht} \left( \partial_{\sigma} x^{i} (t,\sigma) \right)^2
\right]
\simeq 0.
\label{constr-hamiltonian-like}
\end{align}
The equations (\refeq{constr-level-matching}) and (\refeq{constr-hamiltonian-like})
constrain the dynamics of the fields $x^i(t,\sigma)$. By integrating $\mathit{P} (t, \sigma)$ and $\mathit{H} (t, \sigma)$ on $\sigma $ from $[0, 2 \pi]$, we obtain the time-dependent constraints of the model.

In order to find the general solution to the equations of motion (\refeq{eq-motion-x}), we observe that they display a circular symmetry in the spatial variable. Accordingly, we can solve the equation (\refeq{eq-motion-x}) by elementary methods and show that its general solution can be written in terms of the Hankel functions as follows
\begin{align}
x^i (t,\sigma)  & =  x^{i}_{0} +  p^{i}_{0} t 
\\
& +  \frac{i}{2} \sqrt{\frac{\pi }{2H}} 
\sum_{m>0} 
\left[ 
\alpha^{i}_{m} e^{2im \sigma} + \beta^{i}_{m}  e^{- 2im \sigma}
\right] H^{(2)}_{0} (z_m) 
\nonumber
\\
&-
\frac{i}{2}\sqrt{\frac{\pi }{2H}} \sum_{m>0} 
\left[ 
\alpha^{i}_{-m} e^{- 2im \sigma} + \beta^{i}_{-m}  e^{ 2im \sigma}
\right] H^{(1)}_{0} (z_m),
\label{solutions-x} 
\end{align}
where we have used the following notation for the arguments of the Hankel functions 
\begin{equation}
z_m = z_m (t) = \frac{m}{H} e^{2Ht},
\hspace{0.5cm}
m \in \mathbb{N}^{*}.
\label{notation-zm}
\end{equation}
Some remarks are in order here. The integration constants $ x^{i}_{0}$ correspond to the local coordinates of the centre of the circular section of the image of $\Sigma^{1,1}$ under the embedding $x$ at any given time. The integration constants $p^{i}_{0}$ are associated with the variables $x^{i}_{0}$ and describe the motion of the centre of the circle. These constants are analogous to the coordinates and momenta of the centre-of-mass of the closed string in the Minkowski space-time, hence the notation. The constants $\alpha^{i}_{m}$ and $\beta^{i}_{m}$ are the coefficients of the linearly independent solutions in the superposition (\refeq{solutions-x}). They represent the excitations of the $\sigma$-model fields of a definite mode $m$ and along the $x^i$-th coordinate of the de Sitter space. On the two-dimensional base space, the discrete index $m$ can be associated with the two-dimensional momentum. Note that the time-independent  excitations are specific to the cosmological gauge. In an arbitrary time-dependent gauge, the modes are necessarily time-dependent, too. Moreover, the general solution cannot be decomposed in terms of the Hankel functions and, for that matter of fact, in terms of any known functions (see e. g. \cite{Li:2007gf}). The general solution given by the equations (\refeq{solutions-x}) reflect the local geometry of the base space and of the target space. It has similar properties with the solutions obtained in the study of the bosonic string near singularities of the null cosmology from \cite{Madhu:2009jh}. 

Let us now turn our attention to the Hamiltonian formulation of the model. By a simple inspection of the Lagrangian (\refeq{action-cosmo-gauge}), one can see that the Hessian of the time-derivatives of fields is non-singular. Therefore, one can define the canonically conjugate momenta $\pi^{i}(t,\sigma)$ associated to the fields $x^{i}(t,\sigma)$ by the following relation
\begin{equation}
\pi^{i} (t,\sigma)= \frac{1}{2 \pi} \partial_{t} x^{i} (t,\sigma).
\label{pi-def}
\end{equation}
By using the equations (\refeq{solutions-x}) and (\refeq{pi-def}), respectively, we can show that the conjugate momenta associated to the general solution has the form
\begin{align}
\pi^i (t,\sigma) & =  \frac{1}{2 \pi}  p^{i}_{0}  
 + \frac{i}{2}\sqrt{\frac{H}{2 \pi}} \sum_{m>0} 
\left[ 
\alpha^{i}_{m} e^{2im \sigma} + \beta^{i}_{m}  e^{- 2im \sigma}
\right] z_m \frac{d H^{(2)}_{0} (z_m)}{d z_m } 
\nonumber
\\
&-
\frac{i}{2}\sqrt{\frac{H}{2 \pi}} \sum_{m>0} 
\left[ 
\alpha^{i}_{-m} e^{- 2im \sigma} + \beta^{i}_{-m}  e^{ 2im \sigma}
\right] z_m \frac{d H^{(1)}_{0} (z_m)}{d z_m } . 
\label{solutions-pi} 
\end{align}
The formulation of the dynamics of the $\sigma$-model in the phase space defined by the field variables $x^i(t,\sigma)$ and $\pi^i(t,\sigma)$ is straightforward.
The Legendre transformation of the Lagrangian density given by the equation (\refeq{action-cosmo-gauge}) produces the following time-dependent canonical Hamiltonian density
\begin{equation}
\mathit{H}_c (t , \sigma) = \frac{1}{4 \pi} 
\left[
e^{-2Ht} +  
4 \pi^2 \left( \pi^{i} (t,\sigma) \right)^2 
+ 
\left( \partial_{\sigma} x^{i} (t,\sigma) \right)^2 
\right].
\label{Hamiltonian-canonical}
\end{equation}
The system is constrained by $\mathit{P} (t, \sigma)$ and $\mathit{H} (t,\sigma)$ that can be readily expressed in terms of the canonical variables. In order to understand their action on the two-dimensional field theory, we need to interpret them from the point of view of the base space. To this end, we calculate their
Poisson brackets with the field variables $x^i(t,\sigma)$ and obtain the following results
\begin{align}
\{ x^{i}(\sigma) , \mathit{P} (\sigma ') \}_{PB}
& = 
\frac{1}{2} \delta (\sigma - \sigma ' ) \frac{\partial x^{i}(\sigma ')}{\partial \sigma '} ,
\label{PB-xP}
\\
\{ x^{i}(\sigma) , \mathit{H} (\sigma ') \}_{PB}
& = 
\pi \delta (\sigma - \sigma ' ) \pi^{i}(\sigma '),
\label{PB-xH}
\end{align}
where $\delta (\sigma )$ is the delta-function of the finite interval $[0, 2 \pi ]$ and the time variable has been omitted for simplicity. The equation (\refeq{PB-xP}) proves that the constraint $\mathit{P} (t, \sigma )$ generates the translation of the fields along the spatial direction of $\Sigma^{1,1}$. Therefore, $\mathit{P} (t, \sigma )$ corresponds to the density of the momentum on the base space. The second equation (\refeq{PB-xH}) shows that the constraint $\mathit{H} (t, \sigma)$ is the density of the energy functional and generates the translations in time (the Hamiltonian density). This interpretation holds only locally on the target space as the time-evolution takes place along a local time-like Killing vector in the de Sitter space. According to the orientation given by the equation (\refeq{static-gauge}), the time-like vector which is normal to the hyperboloid has the integral lines identical with the time-like lines of the base space. 
 
The Poisson brackets between the constraints $\mathit{P} (t, \sigma)$ and $\mathit{H} (t,\sigma)$ and the momenta $\pi^{i}(t, \sigma)$ can be calculated, too, and the results are given by the following relations
\begin{align}
\{ \pi^{i}(\sigma) , \mathit{P} (\sigma ') \}_{PB}
& = 
\frac{1}{2} \frac{\partial \delta (\sigma - \sigma ' )}{\partial \sigma} \pi^{i} ( \sigma ' ) ,
\label{PB-pP}
\\
\{ \pi^{i}(\sigma) , \mathit{H} (\sigma ') \}_{PB}
& = 
\frac{e^{2Ht}}{4 \pi} \frac{\partial \delta (\sigma - \sigma ' ) }{\partial \sigma '}  \frac{\partial x^{i}(\sigma ')}{\partial \sigma '}.
\label{PB-pH}
\end{align}
The last equation above shows that the temporal translations generated by the Hamiltonian density do not conserve the energy due to a factor that comes from the interaction between the fields with the background metric. The first equation above confirms the interpretation of the momentum as the generator of spatial translations on the base space. By inspecting the equations (\refeq{constr-hamiltonian-like}) and (\refeq{Hamiltonian-canonical}), respectively, one can see that the canonical Hamiltonian density is identical with the Hamiltonian density up to a time-dependent function. This implies that the canonical Hamiltonian does not vanish on the constraint surface at every value of time which is in agreement with the general behaviour of the time-dependent Hamiltonians. 

The formal argument given above can be extended to the arbitrary time-dependent gauges $\omega (t)$ and thus we can conclude that the interpretation of the constraints as two-dimensional densities of momentum and energy should hold in these gauges, too. That is in agreement with the point of view of \cite{Li:2007gf}. Finally, we note that the general solution given by the equations (\refeq{solutions-x}) and (\refeq{solutions-pi}), respectively, are gauge and frame dependent. They add to the body of solutions obtained in the literature in other gauges. Despite this limitation, they are interesting as the theory described by them can be quantized by applying canonical methods.

\section{Canonical Quantization}

In this section, our primary interest is to study the quantization of the 
$\sigma$-model obtained above. The fact that the fields $x^i(t, \sigma)$ and 
$\pi^i(t, \sigma)$ have a decomposition in terms of oscillating modes and the expansion of $\mathit{P} (t,\sigma)$ and $\mathit{H} (t,\sigma)$ in term of oscillators suggest that the model is suitable to the Gupta-Bleuler quantization method. That can be done in a fairly straightforward way with the benefit that the canonical formalism provides the natural framework to discuss the entanglement which is our main goal.

We start as usual by promoting the dynamical fields $x^i(t, \sigma)$ and 
$\pi^i(t, \sigma)$ to operators on the space of states of the $\sigma$-model. 
The requirement that the fields $x^{i}(t, \sigma)$ satisfy the reality condition
implies that $\alpha^{i}_{-m} = \alpha^{i \dagger}_{m}$ and $\beta^{i}_{-m} = \beta^{i \dagger}_{m}$ for all directions and all modes $m > 0$. These are the standard operators that belong to two copies of the harmonic oscillator algebra for each mode. Therefore, they must satisfy the following commutation relations 
\begin{equation}
\left[
\alpha^{i}_{m}, \alpha^{j \dagger}_{n}
\right] =
\left[
\beta^{i}_{m}, \beta^{j \dagger}_{n}
\right] = \delta^{ij} \delta_{mn},
\hspace{0.5cm}
\left[
\alpha^{i}_{m}, \beta^{j}_{n}
\right] = 0,
\label{ab-comm-rel}
\end{equation}
for all $i, j =  1, 2, 3$ and all $m > 0$. Also, we require that the field operators $x^i(t, \sigma)$ and $\pi^i(t, \sigma)$ satisfy the equal-time commutation relation on every spatial section of the embedded cylinder
\begin{equation}
\left[ x^{k} (t, \sigma) , \pi^{l}(t, \sigma') \right] = i \delta^{kl} \delta(\sigma - \sigma' ),
\hspace{0.5cm}
\left[ x^{k}_{0} , \pi^{l}_{0} \right] = i \delta^{kl},
\label{ETC}
\end{equation}
where $k,l = 1, 2, 3$. It is important to note that in the first of the commutators from the equation (\refeq{ETC}), the terms from the left hand side depend explicitly on time while the ones from the right hand side are time-independent. That raises the question of compatibility between the general solution of the equations of motion obtained in an arbitrary time-dependent gauge and the equal time commutation relations. As discussed in the Appendix A, the two are compatible in the cosmological gauge.

\subsection{Quantum Constraints}

The operators that determine the quantum dynamics of the $\sigma$-model are 
$\mathit{H}_c (t)$, $\mathit{H} (t)$ and $\mathit{P} (t)$ and they can be obtained from the corresponding classical functions by the usual procedure. They are defined as the integrated operator densities $\mathit{H}_c (t,\sigma)$, $\mathit{H} (t,\sigma)$ and $\mathit{P} (t,\sigma)$, respectively.
After some lengthy calculations in which the equations (\refeq{solutions-x}) and (\refeq{solutions-pi}) interpreted as operatorial relations and the properties of the Hankel functions are used (see e. g. \cite{Olver:2010}), the following expression for the canonical Hamiltonian in terms of the time-independent oscillators is obtained
\begin{equation}
\mathit{H}_c (t) = \frac{(p^{i}_0)^2}{4} + \sum_{m>0}
\left[
\Omega_{m} (z_m) 
\left(
\alpha^{i \dagger}_{m} \alpha^{i}_{m}
+
\beta^{i \dagger}_{m} \beta^{i}_{m}
\right)
+
\Phi^{(1)}_{m} (z_m)  \alpha^{i}_{m} \beta^{i}_{m}
+
\Phi^{(2)}_{m} (z_m)  \alpha^{i \dagger}_{m} \beta^{i \dagger}_{m} 
\right] + h_{0} (t).
\label{quantum-Hamiltonian}
\end{equation}
Here, we have introduced the following shorthand notations
\begin{align}
\Omega_{m} (z_m) & =  \pi H z^{2}_{m} 
\left[
2 \left( \frac{d Y_{0}(z_m)}{d z_m}\right)^{2}
+ \frac{1}{2} \left( Y_{0}(z_m) \right)^{2}
\right],
\label{Omega}
\\
\Phi^{(1)}_{m} (z_m) & =  i\pi H z^{2}_{m} 
\left[
2  \frac{d Y_{0}(z_m)}{d z_m} \frac{d H^{(2)}_{0}(z_m)}{d z_m}
+ 
\frac{1}{2} Y_{0}(z_m) H^{(2)}_{0}(z_m)
\right]
\label{Phi-1}
\\
\Phi^{(2)}_{m} (z_m) & = - i\pi H z^{2}_{m} 
\left[
2  \frac{d Y_{0}(z_m)}{d z_m} \frac{d H^{(1)}_{0}(z_m)}{d z_m}
+ 
\frac{1}{2} Y_{0}(z_m) H^{(1)}_{0}(z_m)
\right]
\label{Phi-2}
\end{align}
where $Y_0 (z)$ is the Bessel function of the second kind. By using the general properties of the Bessel functions, one can prove that these coefficients obey the following reality conditions
\begin{equation}
\Omega_{m} (z_m) = \frac{1}{2}\mathfrak{Re}\left[ \Phi^{(1)}_{m} (z_m) \right],
\hspace{0.5cm}
\Phi^{(1)}_{m} (z_m) = \Phi^{(2) \star}_{m} (z_m),
\label{reality-O-P}
\end{equation}
for all $m > 0$ and all values of time. The last term from the right hand side of the equation (\refeq{quantum-Hamiltonian}) is the canonical Hamiltonian in the absence of any excitation. This is a time-dependent function that expresses the energy of the time-independent oscillator vacua as well as of the de Sitter background represented by the expansion factor
\begin{align}
h_0 (t) = \frac{1}{4} e^{-4Ht} + 
4i\sum_{m > 0}
\left[
Y_{0}(z_m) H^{(2)}_{0}(z_m)
+
\frac{d Y_{0}(z_m)}{d z_m} \frac{d H^{(2)}_{0}(z_m)}{d z_m}
\right]
\label{h-0}
\end{align}
The function $h_0 (t)$ is analogous to the vacuum energy in the Minkowski space-time. However, as can be seen from the equation (\refeq{h-0}), the vacuum energy in the de Sitter space depends on time in a rather complicate way as a result of the interaction between the fields and the background metric. 

Next, we need to write the constraint operators in term of excitations. By using the equation (\refeq{quantum-Hamiltonian}) and the commutation relations (\refeq{ab-comm-rel}), one can show that the Hamiltonian operator $\mathit{H} (t)$ has the following form
\begin{equation}
\mathit{H} (t)  =  \frac{(p^{i}_0)^2}{4} + \sum_{m>0}
\left[
\Omega_{m} (z_m) 
\left(
\alpha^{i \dagger}_{m} \alpha^{i}_{m}
+
\beta^{i \dagger}_{m} \beta^{i}_{m}
\right)
+
\Phi^{(1)}_{m} (z_m)  \alpha^{i}_{m} \beta^{i}_{m}
+
\Phi^{(2)}_{m} (z_m)  \alpha^{i \dagger}_{m} \beta^{i \dagger}_{m} 
\right]
 +  h^{(2)}_{0} (t), 
\label{quantum-constr-hamilt}
\end{equation}
where
\begin{equation}
h^{(2)}_{0} (t) = h_{0} (t) - e^{-2Ht}. 
\label{const-phi-h}
\end{equation}
From the equations (\refeq{quantum-Hamiltonian}) and (\refeq{quantum-constr-hamilt}) we can see that the operators $\mathit{H}(t)$ and $\mathit{H}_c (t)$ differ from each other on the constraint surface by a time-dependent function only. The last operator $\mathit{P} (t)$ can be obtained in the same way. Some simple calculations lead to the following expression
\begin{equation}
\mathit{P} (t)  =  \frac{\pi}{H}\sum_{m > 0}
m
\frac{d J_{0}(z_m)}{d z_m} Y_{0}(z_m)
\left( 
\alpha^{i \dagger }_{m} \alpha^{i}_{m} 
-
\beta^{i \dagger }_{m} \beta^{i}_{m}
\right),
\label{quantum-constr-level}
\end{equation} 
where $J_0 (z) $ is the Bessel function of the first kind. The constraints generated by the operators $\mathit{P} (t)$ and $\mathit{H} (t)$ on the states of the system are the quantum counterpart of the classical equations (\refeq{constr-level-matching}) and (\refeq{constr-hamiltonian-like}), respectively. These constraints can be treated in the Gupta-Bleuler formalism in which they are equivalent with the definitions of the kernels of the operators $\mathit{P} (t)$ and $\mathit{H} (t)$, respectively. Thus, the subspace of the physical states is the intersection of the kernels of the operators $\mathit{P} (t)$ and $\mathit{H} (t)$ at every value of the time parameter. 

\subsection{Time-Independent Hilbert Space}

We will now proceed to describe the states of the model. It follows from the relations (\refeq{ab-comm-rel}) and (\refeq{ETC}) that there is a time-independent Hilbert space $\mathcal{H}$ associated with the quantum $\sigma$-model that can be decomposed into the following direct product
\begin{equation}
\mathcal{H}  = \mathcal{H}_{0} \bigotimes \mathcal{F} 
= 
\mathcal{H}_{0} \bigotimes 
\left[
\mathcal{F}_{\alpha} \bigotimes \mathcal{F}_{\beta}
\right] 
=\mathcal{H}_{0} \bigotimes_{(i,m)} \mathcal{F}_{(i,m)}.
\label{Hilbert-decomposition}
\end{equation} 
Here, $\mathcal{H}_{0}$ is the Hilbert space spanned by the eigenstates of the operators $\hat{p}^{i}_{0}$'s that are solutions of the following equations
\begin{equation}
\hat{p}^{i}_{0} |p^{j}_{0} \rangle   =  \delta_{ij} p^{i}_{0}  |p^{j}_{0}  \rangle  ,
\label{p-state}
\end{equation}
where we have used the hat as a one-time notation for the operators and there is no summation in the right hand side of the equation (\refeq{p-state}). The 
$\mathcal{F}$ factor of the direct product from the equation (\refeq{Hilbert-decomposition}) is the time-independent Fock space of the oscillating field excitations. On its turn, $\mathcal{F}$ can be decomposed into pairs of oscillators from $\alpha$ and $\beta$ sectors, respectively, of the same mode $m$ and in the same direction $i$. The Fock space of each pair is denoted by 
$\mathcal{F}_{(i,m)}$. 

A general time-independent oscillator state from $\mathcal{F}$ can be constructed as usual by acting with the creation operators on the vacuum state of all oscillators defined by the following equations
\begin{equation}
\alpha^{i}_{m} | 0 \rangle  = \beta^{i}_{m} | 0 \rangle = 0,
\label{vacuum-state}
\end{equation}
in all spatial directions $i$ and for all $m>0$. Then the quantization method provides the canonical basis of $\mathcal{F}$ that contains vectors of the form
\begin{equation}
\left( \alpha^{i_1 \dagger}_{m_1} \right)^{p^{i_1}_{m_1}}
\left( \alpha^{i_2 \dagger}_{m_2} \right)^{p^{i_2}_{m_2}}
\cdots
\left( \beta^{j_1 \dagger}_{n_1} \right)^{q^{j_1}_{n_1}}
\left( \beta^{j_2 \dagger}_{n_2} \right)^{q^{j_2}_{n_2}}
\cdots
| 0 \rangle .
\label{basis-Fock-space}
\end{equation}
The above construction is nothing more than the Bargmann-Fock representation of the $\sigma$-model which follows from the equations (\refeq{ab-comm-rel}) and (\refeq{ETC}) that describe a standard quantum field theory on the embedded two-dimensional cylinder. 

In order to designate the time-independent oscillator states, it proves useful to use the multi-index notation
\begin{align}
\mathcal{N}_{\alpha} & =  
\{ N^{i}_{\alpha, m}\}^{i = 1, 2, 3}_{m > 0} =
\{ N^{1}_{\alpha, 1}, N^{2}_{\alpha, 1}, N^{3}_{\alpha, 1};
N^{1}_{\alpha, 2}, N^{2}_{\alpha, 2}, N^{3}_{\alpha, 2};
\cdots
\} ,
\label{multi-index-a}
\\
\mathcal{N}_{\beta} & =  
\{ N^{i}_{\beta, m}\}^{i = 1, 2, 3}_{m > 0} =
\{ N^{1}_{\beta, 1}, N^{2}_{\beta, 1}, N^{3}_{\beta, 1};
N^{1}_{\beta, 2}, N^{2}_{\beta, 2}, N^{3}_{\beta, 2};
\cdots
\} ,
\label{multi-index-b}
\end{align}
Here,  $N^{i}_{\alpha, m}$ and $N^{i}_{\alpha, m}$ are natural numbers and they represent the eigenvalues of the number operators $\hat{N}^{i}_{\alpha, m}$ and $ \hat{N}^{i}_{\beta, m}$, respectively. In this notation, the elements of the canonical basis are labelled by pairs of multi-indices from both sectors 
\begin{equation}
\{ ( \mathcal{N}_{\alpha} , \mathcal{N}_{\beta} ) \}^{i = 1, 2, 3}_{m > 0}\leftrightarrow
\{ | \mathcal{N}_{\alpha} ; \mathcal{N}_{\beta}\rangle \}^{i = 1, 2, 3}_{m > 0} .
\label{multi-index-basis}
\end{equation}
We can use the multi-index notation to represent an arbitrary time-independent oscillator state from $\mathcal{F}$ as
\begin{equation}
| \Psi \rangle_{osc} = \sum_{\mathcal{N}_{\alpha}} \sum_{\mathcal{N}_{\beta}}
C(\mathcal{N}_{\alpha} ; \mathcal{N}_{\beta}) | \mathcal{N}_{\alpha} ; \mathcal{N}_{\beta}\rangle \in \mathcal{F},
\label{multi-general-vector}
\end{equation}
where the multi-sum notation over the multi-indices is understood and $C(\mathcal{N}_{\alpha} ; \mathcal{N}_{\beta})$ are complex numbers.

Now let us examine the kernel of the operators $\mathit{P} (t)$ and $\mathit{H} (t)$ in the time-independent Hilbert space $\mathcal{H}$. We see from the equation (\refeq{quantum-constr-level}) that the constraint generated by the two-dimensional momentum operator
\begin{equation}
\mathit{P} (t) | \Psi \rangle  = 0,
\label{phys-state-1}
\end{equation}
selects those states from $\mathcal{H}$ with an equal number of $\alpha$ and $\beta$ excitations at any value of time. Therefore, the constraint (\refeq{phys-state-1}) is analogous to the familiar level matching condition in the Minkowski space-time. Beside its action on the oscillator Fock space, the operator $\mathit{P} (t)$ acts multiplicatively on the complex coefficients $C(\mathcal{N}_{\alpha} ; \mathcal{N}_{\beta})$ and turns them into time-dependent functions. 

The kernel of the  $\mathcal{H}$ is defined by the following equation
\begin{equation}
\mathit{H} (t) | \Psi \rangle  = 0.
\label{phys-state-2}
\end{equation}
It follows from the equation (\refeq{quantum-constr-hamilt}) that the operator $\mathit{H}(t)$ acts on all factors of the direct product of $\mathcal{H}$. However, 
the equation (\refeq{phys-state-2}) does not admit any solution in the full time-independent Hilbert space since the Hamiltonian operator has eigenvectors just in the $\mathcal{H}_{0}$ subspace. The time-independent oscillator states are in fact quantum quenches of $\mathit{H} (t)$ that generates the entanglement of the oscillators from the left and right sectors through its off-diagonal terms. Also, it acts on the coefficients  $C(\mathcal{N}_{\alpha} ; \mathcal{N}_{\beta})$ by multiplying them with time-dependent functions. An immediate consequence of that is the fact that the equation (\refeq{phys-state-2}) cannot be used to determine the mass or energy of the states that belong to $\mathcal{H}$ as opposed to the Minkowski space-time.  

\subsection{Time-Dependent Hilbert Space}

The above analysis suggests that the time-independent vectors from  $\mathcal{H}$ should be tensored with time-dependent functions in order to describe the states of the $\sigma$-model. Another argument that supports this construction is provided by the field excitations: while the oscillator operators are time-independent, their frequencies are not. There is a priori no restriction on the time-dependent smooth functions to be used. However, the requirement that the states be physical results on constraints on these functions that 
arise from the action of the operator $\mathit{H} (t)$. 

In order to determine these constraints, consider the action of the Hamiltonian operator on a general time-independent state of the form $|\Psi \rangle = \sum_{i} | p^{i}_{0} \rangle | \Psi \rangle_{osc}$. A short calculation that involves the equations (\refeq{quantum-constr-hamilt}) and (\refeq{multi-general-vector}), respectively, shows that $|\Psi \rangle$  is in the kernel of $\mathit{H} (t)$ if the following equation holds
\begin{align}
&\left[ 
\frac{(p^{i}_0)^2}{4} + h^{(2)}_{0} (t)
+ \sum^{3}_{i=1} \sum_{m > 0} \Omega_{m} (z_m) 
\left( 
N^{i}_{\alpha, m} + N^{i}_{\beta, m}
\right)
\right]
C(\mathcal{N}_{\alpha} ; \mathcal{N}_{\beta})
\nonumber
\\
&+ \sum^{3}_{i=1} \sum_{m > 0} \Phi^{(1)}_{m} (z_m)
\sqrt{(N^{i}_{\alpha, m} + 1)(N^{i}_{\beta, m} + 1)}
C(\check{\mathcal{N}}_{\alpha} ; \check{\mathcal{N}}_{\beta};N^{i}_{\alpha, m} + 1,N^{i}_{\beta, m} + 1 )
\nonumber
\\
&+\sum^{3}_{i=1} \sum_{m > 0} \Phi^{(2)}_{m} (z_m)
\sqrt{N^{i}_{\alpha, m} N^{i}_{\beta, m} }
C(\check{\mathcal{N}}_{\alpha} ; \check{\mathcal{N}}_{\beta};N^{i}_{\alpha, m} - 1,N^{i}_{\beta, m} - 1 ) = 0.
\label{ground-state-phys-indep-no-eq}
\end{align}
Here, the multi-index checked means that $N^{i}_{m}$ corresponding to the indices $(i,m)$ is replaced by $N^{i}_{m} \pm 1$. The equation (\refeq{ground-state-phys-indep-no-eq}) is a recurrence relation among the coefficients $C(\mathcal{N}_{\alpha} ; \mathcal{N}_{\beta})$. However, the factors that enter this equation are functions of time which implies that the coefficients $C(\mathcal{N}_{\alpha} ; \mathcal{N}_{\beta})$ must depend on time, too. 

These arguments show that, in order to be able to describe the physical states of the system, one has to generalized the time-independent Hilbert space to a one-parameter family of Hilbert spaces  parametrized by the local time. This family is obtained by tensoring $\mathcal{H}$ with smooth functions on time in the north diamond of the Penrose diagram
\begin{equation}
\mathbf{H} (\mathbb{R}^{+}) = \{ \mathcal{H}_t \}_{t \in \mathbb{R}^{+}}
= \mathsf{F} (\mathbb{R}^{+}) \bigotimes \mathcal{H}.
\label{Hilbert-space-time}
\end{equation}
A general state of the system at a given value of time $t$ is a linear superposition of the elements of the time-independent basis of $\mathcal{H}$ with coefficients from $\mathsf{F}(t)$ that represent the values of the smooth functions at that given time.

The physical states form a subspace of $\mathbf{H} (\mathbb{R}^{+})$ that can be characterized as follows. The action of the operators $\mathit{P} (t)$ and $\mathit{H} (t)$ on the space $\mathbf{H} (\mathbb{R}^{+})$ is natural: they act multiplicatively on the time-dependent factor $\mathsf{F} (\mathbb{R}^{+})$ and 
through oscillator operators on the time-independent space $\mathcal{H}$. (This action can be generalized to any other operator of the model.) The kernel $\mbox{ker}P[\mathbf{H} (\mathbb{R}^{+})]$ of $\mathit{P} (t)$ contains the states with equal number of $\alpha$ and $\beta$ time-independent excitations. The kernel $\mbox{ker}H[\mathbf{H} (\mathbb{R}^{+})]$ of $\mathit{H} (t)$ is defined by the states of which coefficients satisfy the recurrence relation (\refeq{ground-state-phys-indep-no-eq}). The physical states form a subspace of $\mathbf{H} (\mathbb{R}^{+})$ given by the intersection of the two kernels
\begin{equation}
\mathbf{H}_{phys} (\mathbb{R}^{+}) = \mbox{ker}H[\mathbf{H} (\mathbb{R}^{+})]
\bigcap \mbox{ker}P[\mathbf{H} (\mathbb{R}^{+})].
\label{physical-H-family}
\end{equation}

One can get more insight into the structure of the space of states if we look at some examples. The simplest state is the ground state $| \Psi_0 (t) \rangle = | \Psi_0 \rangle$ characterized by the quantum numbers $p^{i}_{0} = 0, N^{i}_{\alpha, m} = N^{i}_{\beta, m} = 0 $ for all pairs $(i,m)$. This state is time-independent and belongs to $\mbox{ker}P[\mathbf{H} (\mathbb{R}^{+})]$ by definition. The action of the Hamiltonian on it produces the following equation
\begin{equation}
\sum^{3}_{i=1} \sum_{m > 0} \Phi^{(2)}_{m} (z_m)|1^{i}_{\alpha, m}; 1^{i}_{\beta, m} \rangle + h^{(2)}_{0} (t) | 0 \rangle = 0.
\label{example-ground-state}
\end{equation}
The equation (\refeq{example-ground-state}) does not have any solution since the vectors are linearly independent and their coefficients do not vanish for an arbitrary value of time . Thus, the state $| \Psi_0 \rangle $ is not in the kernel of $\mathit{H} (t)$. 

The next simple states are specified by the eigenvalues $p^{i}_{0} \neq 0$ and a linear combination of excitations $N^{i}_{\alpha, m} = N^{i}_{\beta, m} = N$ of a single oscillating mode and in just one direction
\begin{equation}
| \Psi_{1} (t) \rangle = \sum^{\infty}_{N = 0} C_N (t) | N;N \rangle,
\label{example-one-oscillator}
\end{equation}
where we have used the notation $C(N;N;t) = C_N (t)$ and we have dropped the fixed indices $i$ and $m$, respectively. By construction, this linear superposition is in the kernel of the momentum. The action of the Hamiltonian on the state $| \Psi_{1} (t) \rangle$ generates the following recurrence relation
\begin{equation}
\left[ 
2 N \Omega_{m} (z_m) + \frac{(p^{i}_0)^2}{4} + h^{(2)}_{0} (t)
\right] C_N (t) + (N + 1)\Phi^{(1)}_{m} (z_m) C_{N+1} (t)
+ N \Phi^{(2)}_{m} (z_m) C_{N-1} (t) = 0.
\label{recurrence-eq-one-osc}
\end{equation}
The equation (\refeq{recurrence-eq-one-osc}) shows that $| \Psi_{1} (t) \rangle$ actually describes a set of equivalent states up to a time-dependent phase factor that belong to the kernel of the Hamiltonian. Each state is determined by the lowest coefficient  $ C_0 (t)$ which is an arbitrary function on time. Indeed, the equation (\refeq{recurrence-eq-one-osc}) gives for the coefficients the following expressions  
\begin{align}
C_1 (t)& =  - \frac{\Phi^{(1)}_{m} (z_m)}{\frac{(p^{i}_0)^2}{4} + h^{(2)}_{0} (t)} C_0 (t) ,
\nonumber
\\
C_2 (t)& =  - \frac{1}{2 \Phi^{(1)}_{m} (z_m)} 
\left[
\left(
2 \Omega_{m} (z_m) +\frac{(p^{i}_0)^2}{4} + h^{(2)}_{0} (t)
\right)\frac{\Phi^{(1)}_{m} (z_m)}{\frac{(p^{i}_0)^2}{4} + h^{(2)}_{0} (t)}
-
\Phi^{(2)}_{m} (z_m)
\right]
 C_0 (t) ,
\nonumber
\\
& \vdots 
\label{coeff-recurr-example}
\end{align}
The above example illustrates the content of the recurrence relation (\refeq{ground-state-phys-indep-no-eq}) and proves that the kernel of $\mathit{H} (t)$ and the physical subspace that are defined by this relation are not empty sets.

We conclude that, by extending the time-independent Hilbert space to the one-parameter family of spaces $\mathbf{H} (\mathbb{R}^{+})$, we are able to solve the kernel equation of the Hamiltonian operator and to characterize the structure of the physical states of the $\sigma$-model. That is possible because the problem of determining the eigenstates of the Hamiltonian 
becomes the soluble problem of determining the time-dependent smooth coefficients that satisfy the recurrence relation (\refeq{ground-state-phys-indep-no-eq}) in $\mathbf{H} (\mathbb{R}^{+})$. 

\subsection{The Jordan-Schwinger Representation}

In the discussion so far, we have represented the time-independent Hilbert space of the $\sigma$-model oscillators as a product of the time-independent Bargmann-Fock (BF) representations of individual modes. Alternatively, we can organize the Hilbert space to reflect the symmetries of the Hamiltonian, namely in terms of a direct product of irreducible representations of the $SU(1,1)$ group by using the Jordan-Schwinger  (JS) representation \cite{Jordan:1935,Schwinger:1965}. 

Indeed, by inspecting the equation (\refeq{quantum-constr-hamilt}), one recognizes that the operator $\mathit{H} (t)$ has an $SU(1,1)^{\infty}$ structure: each copy of $SU(1,1)$ is associated to a pair of oscillators $\alpha$ and $\beta$, respectively, that have the same $(i,m)$ indices. The Hamiltonian $\mathit{H} (t)$ is a linear combination with time-dependent coefficients of the generators of the $su(1,1)$-algebra in the JS representation
\begin{equation}
K^{i}_{0,m} = \frac{1}{2} \left( \alpha^{i \dagger}_{m} \alpha^{i}_{m} 
+ \beta^{i \dagger}_{m} \beta^{i}_{m} + 1 \right),
\hspace{0.5cm}
K^{i}_{+,m} = \alpha^{i \dagger}_{m} \beta^{i \dagger}_{m},
\hspace{0.5cm}
K^{i}_{-,m} = \alpha^{i}_{m} \beta^{i}_{m}.
\label{Schwinger-generators}
\end{equation} 
The $su(1,1)$ algebras are independent of each other. Thus, the commutators of their generators satisfy the following relations
\begin{equation}
[ K^{i}_{0,m} , K^{j}_{ \pm,n} ] = \pm \delta^{ij} \delta_{mn} K^{j}_{\pm,n},
\hspace{0.5cm}
[ K^{i}_{+,m}, K^{j}_{-,n}] = - 2 \delta^{ij} \delta_{mn} K^{j}_{0,n}. 
\label{su11-algebra}
\end{equation}

There is a map between the BF and JS representations, respectively, by which the bi-oscillator states can be expressed in terms of products of irreducible representations of the $SU(1,1)$ (see e. g. \cite{Perelomov:1986tf})\footnote{We recall the well known fact that the irreducible representations of the $SU(1,1)$ are infinite dimensional as the group is non-compact \cite{Bargmann:1946me}.}. Let us describe it for the system at hand\footnote{For a similar treatment of the two-oscillator representation but in a different context see \cite{Shaterzadeh:2008}.}.  

The irreducible representations of the $su(1,1)$ algebra are classified by the eigenvalues of the operators $K_{0,m}$ and the Casimir operators $C^{i}_{m}$ that are solution to the following eigenvector and eigenvalue equations 
\begin{align}
K^{i}_{0,m} | k^{i}_{m} , \lambda^{i}_{m} \rangle & =  \lambda^{i}_{m} | k^{i}_{m} , \lambda^{i}_{m} \rangle , 
\label{k0}
\\
C^{i}_{m} 
| k^{i}_{m} , \lambda^{i}_{m} \rangle & =  k^{i}_{m} (k^{i}_m -1 )| k^{i}_{m} , \lambda^{i}_{m} \rangle,
\label{k2}
\end{align}
where the Casimir operators are defined by the relation
\begin{equation}
C^{i}_{m}  = \frac{1}{4} \left[ 
\left( \alpha^{i \dagger}_{m} \alpha^{i}_{m} 
- \beta^{i \dagger}_{m} \beta^{i}_{m}\right)^{2} - 1
\right].
\label{Casimir-su11}
\end{equation}
Consider the discrete irreducible representations 
$\mathcal{D}^{\pm}_{k^{i}_{m}}$ of the group $SU(1,1)$ defined by the values
$k^{i}_{m} = \frac{1}{2}, 1, \frac{3}{2}, \ldots$ and $\lambda^{i}_{m} = k^{i}_{m} \pm l^{i}_{m}$, where $l^{i}_{m} = 0, 1, 2, \ldots$ for all $i$ and $m$.
Then the map between the BF and JS representations, respectively, takes the oscillator state labelled by the eigenvalues of the number operators $(N^{i}_{\alpha ,m}, N^{i}_{\beta ,m})$ into the state labelled by the pair $ (k^{i}_{m} , \lambda^{i}_{m} )$. The relation between $k^{i}_{m, \pm}$ and $\lambda^{i}_{m}$ implies that $l^{i}_{m, +} = N^{i}_{\beta ,m}$ and $l^{i}_{m, -} = N^{i}_{\alpha ,m}$, respectively. A short computation shows that these numbers satisfy the following relations
\begin{equation}
k^{i}_{m, \pm} =  
\frac{\pm \left( N^{i}_{\alpha ,m} - N^{i}_{\beta ,m} \right) + 1}{2},
\hspace{0.5cm}
\lambda^{i}_{m} = \frac{N^{i}_{\alpha ,m} + N^{i}_{\beta ,m} + 1}{2}.
\label{km-NN-relation}
\end{equation}
The interpretation of this map is that a bi-oscillator state $|N^{i}_{\alpha ,m} , N^{i}_{\beta ,m} \rangle$ in the BF representation is equivalent with the state $|k^{i}_{m}, \lambda^{i}_{m} \rangle$ in the JS-representation if the quantum numbers satisfy the relations (\refeq{km-NN-relation}). 

As we have seen in the previous subsections, the space of the time-independent oscillator states has the following structure in the BF representation
\begin{equation}
\mathcal{F} = \bigoplus_{i,m} \mathcal{F}^{i}_{m}
 = \bigoplus_{i,m} \left[ \mathcal{F}^{i}_{\alpha, m} \bigotimes \mathcal{F}^{i}_{\beta, m} \right],
\label{Fock-ab-time-indep}
\end{equation}
Then the map given by the equation (\refeq{km-NN-relation}) implies that the subspace corresponding to a given pair $(i,m)$ has the following decomposition in terms of discrete irreducible representations of the $SU(1,1)$ group 
\begin{equation}
\mathcal{F}^{i}_{m} =  \mathcal{D}^{+}_{\frac{1}{2}} 
\bigoplus_{s^{i}_{m}}
\left[ 
\mathcal{D}^{+}_{s^{i}_{m} + \frac{1}{2} } \bigoplus \mathcal{D}^{-}_{s^{i}_{m} + \frac{1}{2}}
\right],
\label{Fock-space}
\end{equation}
where $s^{i}_{m} = \frac{1}{2}, 1, \frac{3}{2}, 2, \ldots$. We observe that the Hamiltonian $\mathit{H} (t)$ is invariant under a second set of symmetries that transform simultaneously the left and right modes into one another 
\begin{equation}
\alpha^{i }_{m} \leftrightarrow \beta^{i}_{m} \,  ,
\hspace{0.5cm} \forall m > 0, \, \forall i = 1, 2, 3 \, .
\label{a-b-inversion}
\end{equation}
The transformations (\refeq{a-b-inversion}) induces an equivalence relation between the representations $\mathcal{D}^{+}_{s^{i}_{m}} \simeq \mathcal{D}^{-}_{s^{i}_{m}}$. This gives the final decomposition of the Fock space in terms of either positive or negative discrete irreducible representations of the $SU(1,1)$ group as
\begin{equation}
\mathcal{F} = \bigoplus_{i,m} \mathcal{F}^{i}_{m}
 = \bigoplus_{i,m} \left[ \mathcal{D}^{+}_{\frac{1}{2}} 
\bigoplus_{s^{i}_{m}}
2 \mathcal{D}^{+}_{s^{i}_{m} + \frac{1}{2}} \right].
\label{Fock-ab-time-indep-final}
\end{equation}

We conclude this section by describing the physical subspace in the JS representation. From the previous sections, we know that it is only the momentum operator that restricts the time-independent oscillator states, therefore we need to study just the kernel of $\mathit{P} (t)$. The time-dependent coefficients of the physical states are independent of the representation adopted to describe the excitations. Therefore, the $\mbox{ker}H[\mathbf{H} (\mathbb{R}^{+})]$ is the same as before and the time-dependent coefficients of the physical states are determined by the recurrence relation (\refeq{ground-state-phys-indep-no-eq}) among the coefficients. The states from $\mbox{ker}P[\mathbf{H} (\mathbb{R}^{+})]$ that were given by the level matching condition in the BF representation are mapped into physically equivalent states of the JS representation by the map from the equation (\refeq{km-NN-relation}) if the following equations are satisfied 
\begin{align}
k^{i}_{m} = \frac{1}{2}, \hspace{0.5cm}
\mu^{i}_{m} = N^{i}_{m} + \frac{1}{2} ,
\label{physical-su11}
\\
N^{i}_{m} = N^{i}_{\alpha ,m} = N^{i}_{\beta ,m},
\label{notation-N}
\end{align}
for all $i = 1, 2, 3$ and all $m > 0$. Thus, the time-dependent physical subspace  is
\begin{equation}
\mathbf{H}_{phys} (\mathbb{R}^{+}) = \mbox{ker}H[\mathbf{H} (\mathbb{R}^{+})]
\bigcap [ 
\mathsf{F} (\mathbb{R}^{+}) \bigoplus_{i,m} \mathcal{D}^{+}_{\frac{1}{2}} ] .
\label{physical-H-family-1}
\end{equation}
The above sum contains infinitely many identical terms corresponding to all bi-oscillators $\alpha - \beta$. 

\section{Time-Dependent Entanglement Entropy}

Having discussed the one-parameter family of Hilbert spaces, we turn now our attention to the entanglement between the $\alpha$ and $\beta$ modes generated by the de Sitter background. One good measure of it is the LREE of the ground state of the model \cite{Nielsen:2011}. However, since the Hamiltonian acts on the ground state and modifies the entanglement, the relevant quantity that should be calculated is the entanglement entropy of the ground state acted upon by the Hamiltonian during a finite time interval.

\subsection{Partial Evolution Map}

From the construction done in the previous section, we know that 
$\mathit{H}(t)$ generates a map $\Upsilon (t_2,t_1)$ between two arbitrary sections of the one-parameter family $\mathcal{H}_{t_1}$ and $\mathcal{H}_{t_2}$, respectively. This map does not describe the time-evolution of the total system which includes the de Sitter background beside the $\sigma$-model. In order to see how $\Upsilon (t_2,t_1)$ is related with the time-evolution of the system, we consider that the relevant scale of the embedded base space $x(\Sigma^{1,1})  \subset dS_4$ is much smaller than that of the de Sitter space. Then, by invoking the principle of equivalence in a small neighbourhood of $x(\Sigma^{1,1})$, the total Hamiltonian can be decomposed as follows   
\begin{equation}
\mathit{H}_{tot} (t) = \mathit{H} (t) + \mathit{H}_{B} (t) + \mathit{H}_{int} (t),
\label{Hamiltonian-total}
\end{equation}
where $\mathit{H}_B (t)$ is the Hamiltonian of the background degrees of freedom and  
$\mathit{H}_{int} $ is the Hamiltonian that describes quantum interactions between the $\sigma$-model and the background other than the entanglement of the left and right modes and the generation of time-dependent frequencies (interactions with the classical background). It is clear that the equation (\refeq{Hamiltonian-total}) is an approximation of the unknown Quantum Gravity to a local Quantum Field Theory. For short time intervals, it is reasonable to  approximate further the total Hamiltonian $\mathit{H}_{tot} (t)$ by considering a constant interaction Hamiltonian as well as a constant $\mathit{H}_{B} (t)$. It follows that an arbitrary state of the total system can be represented schematically by the following density matrix
\begin{equation}
\rho_{tot} (t) = \sum_{\sigma' , \sigma} \sum_{B',B} c^{\star}_{\sigma' B'} (t)
c_{\sigma B}  (t)
|\Psi_{\sigma'} (t) \rangle |\Psi_{B'} (t) \rangle 
\langle \Psi_{B} (t) | \langle \Psi_{\sigma} (t) | \,
\label{density-matrix-total}
\end{equation} 
where $|\Psi_{\sigma} (t) \rangle$ and $|\Psi_{B} (t)  \rangle$ denote the vectors of the orthonormal basis of the $\sigma$-model and the de Sitter background, respectively. Since the total system is closed, the time-evolution of 
$\rho_{tot} (t)$ is given by the following equation
\begin{equation}
\rho_{tot} (t_2) = U_{tot} (t_2 , t_1) \rho_{tot} (t_1) U^{\dagger}_{tot} (t_2 , t_1)
\label{density-matrix-total-evol}
\end{equation}
where the time-evolution operator is 
\begin{equation}
U_{tot} (t_2 , t_1 ) = \mathcal{T} \exp 
\left[
- i \int^{t_2}_{t_1} dt \mathit{H}_{tot} (t)
\right] .
\label{evol-op-total}
\end{equation}

In order to separate that component $\rho_{\sigma} (t)$ of $\rho_{tot} (t)$ that contains  information about the entanglement of the left and right modes, one has to trace out the background degrees of freedom. In the case of the total states that contain the $\sigma$-model ground state $|\Psi_{0} (t) \rangle$ one can calculate this trace and show that  
\begin{equation}
\rho_{\sigma, \Psi_0} (t) = f_{0} (t) \, \rho_{\Psi_0} (t), 
\label{density-matrix-sigma-bkg}
\end{equation} 
where 
\begin{equation}
\rho_{\Psi_0} (t) = |\Psi_{0} (t) \rangle \langle \Psi_{0} (t) | \, ,
\hspace{0.5cm}
f_{0} (t) = \sum_{B} |c_{0 B}  (t)|^{2}.
\label{density-matrix-gst-sigma}
\end{equation}
If we make a further simplifying assumption that $\mathit{H}_{int} \simeq 0$ for the time interval considered, then the operator $\Upsilon (t_2,t_1)$ is represented by
\begin{equation}
\Upsilon (t_2,t_1) = \mathcal{T}
\left[ 
\exp\left(-i 
	\int^{t_2}_{t_1} dt  \mathit{H}(t)
	 \right)
\right].
\label{evolution-operator}
\end{equation}
This is a partial evolution operator that acts only on the $\sigma$-model modes and in general is not unitary. A similar partial evolution operator can be defined for the background by replacing the $\mathit{H}(t)$ with the corresponding (unknown) Hamiltonian $\mathit{H}_{B}(t)$. 

On the other hand, the inspection of the equations (\refeq{quantum-constr-hamilt}) and (\refeq{reality-O-P}), respectively, reveals that the non-Hermitian part of $\mathit{H}(t)$ is the complex function $h^{(2)}_{0} (t)$ that contains contributions from the vacuum energy. One can easily verify that the normal ordering procedure renders this energy real\footnote{Actually, the normal ordered operator $\normord{\mathit{H}(t)}$ can be used to define all objects constructed in the previous section such as: the recurrence equation (\refeq{ground-state-phys-indep-no-eq}), the one-parameter family of states given by the equation (\refeq{Hilbert-space-time}) or its subset of physical states from the equation (\refeq{physical-H-family}). Formally, the relations based on $\mathit{H}(t)$ differs from the ones constructed from $\normord{\mathit{H}(t)}$ in that the function $h^{(2)}_{0} (t)$ gets replaced by 
$\normord{h^{(2)}_{0} (t)}$.}
which makes the Hamiltonian Hermitian
\begin{equation}
\normord{\mathit{H}(t)}^{\dagger}  =  \normord{\mathit{H}(t)} \, ,
\hspace{0.5cm}
\normord{h^{(2)}_{0} (t)}  =  \frac{1}{4}e^{4Ht}- e^{-2Ht}, 
\label{normal-ordered-H-h}
\end{equation}  
Thus, the Hamiltonian generated map has the following form
\begin{equation}
\normord{\Upsilon (t_2,t_1)}
 =  \mathcal{T}
\left[ 
\exp\left(-i 
	\int^{t_2}_{t_1} dt \normord{ \mathit{H}(t)}
	 \right)
\right]. 
\label{unitary-map}
\end{equation}
In terms of vector states from the time-dependent Hilbert space, the component of 
$| \Psi (t_1) \rangle$ that belongs to $\mathcal{H}_{t_1} \subset \mathbf{H}_{phys} (\mathbb{R}^{+})$
has its oscillator modes entangled by the non-diagonal terms of the Hamiltonian. This is the most important piece of information about the mapping generated by 
$\normord{\Upsilon (t_2,t_1)}$ as the rest of the Hamiltonian corresponds to free oscillators with time-dependent frequencies.  

\subsection{Entanglement Entropy}

Before proceeding to the technical arguments, let us recall the statement of the problem. We want to determine the entanglement entropy of the $\alpha - \beta$ modes, that is the LREE. To this end we consider only the ground state defined at $t_1$ by the density matrix $\rho_{\Psi_0} (t_1)$ since the change in the   function $f_{0} (t)$ from the equation (\refeq{density-matrix-gst-sigma}) is
\begin{equation}
f(t_2,t_1) = \sum_{B',B} \sum_{B''} \Xi_{B'' B} (t_2 - t_1) \Xi_{B' B''} (t_1 - t_2) \, ,
\hspace{0.5cm}
 \Xi_{B B'} (t_2 - t_1) = \langle \Psi_B (t_2) | U_{B} (t_2 , t_1 ) | \Psi_{B'} (t_1) \rangle.
\label{f-changed}
\end{equation}
The operator $\normord{\Upsilon (t_2,t_1)}$ acts on the state $\rho_{\Psi_0} (t_1)$ at $t_1$ and maps it into the state $\rho_{\Psi_0} (t_2)$ at $t_2$. The entanglement between the left and right modes produced during this process can be computed in the reduced density matrix formalism \cite{Nielsen:2011}, that is given by the trace over either $\alpha$ or $\beta$ degrees of freedom. The corresponding reduced density matrices are $\rho_{\Psi', \beta} (t_1 , t_2)$ and $\rho_{\Psi', \alpha} (t_1 , t_2)$, respectively. 

The first thing to note is that it is advantageous to write $\Upsilon (t_2,t_1)$ in the Jordan-Schwinger representation. After some simple algebra that involves the equations (\refeq{Schwinger-generators}), we obtain
\begin{equation}
\normord{\Upsilon (t_2,t_1)} =  e^{  \frac{i}{2} \Theta(t_1,t_2)}
e^{ - i \sum\limits_{i,m} \left[
2 \omega_{m} (t_1 , t_2 ) K^{i}_{0,m} + 
\phi^{(1)}_{m} (t_1 , t_2 ) K^{i}_{-,m} +
\phi^{(2)}_{m} (t_1 , t_2 ) K^{i}_{+,m}
\right]
},	
\label{evolution-Jordan-Schwinger}
\end{equation}
where $\omega_{m} (t_1 , t_2 )$, $\phi^{(1)}_{m} (t_1 , t_2 )$ and 
$\phi^{(2)}_{m} (t_1 , t_2 )$ are the functions $-i \Omega_{m} (t)$, $-i\Phi^{(1)}_{m} (t)$ and $-i\Phi^{(2)}_{m} (t)$ integrated. The phase factor is given by the following relation
\begin{equation}
\Theta(t_1,t_2) = 2 \sum_{i,m} \omega_{m} (t_1 , t_2 ) 
+ \frac{1}{8H}e^{-4H(t_2 - t_1)} - \frac{1}{H} e^{-2H(t_2 - t_1)}
- \frac{(p^{i}_{0})^{2}}{2}(t_2 - t_1 ).
\label{Theta}
\end{equation}
The operator (\refeq{evolution-Jordan-Schwinger}) can be disentangled in the $su(1,1)$ algebra by using the Baker-Campbell-Hausdorff (BCH) formula \cite{Wilcox:1967zz,Gilmore:1974,Truax:1985vk}. The result is given by the following equation
\begin{equation}
\normord{\Upsilon (t_2,t_1)} =  e^{  \frac{i}{2} \Theta(t_1,t_2)}
\prod_{i,m}
\exp\left[ \chi_{+,m} (t_1 , t_2 ) K^{i}_{+,m} \right]
\exp\left[ \chi_{0,m} (t_1 , t_2 ) K^{i}_{0,m} \right]
\exp\left[ \chi_{-,m} (t_1 , t_2 ) K^{i}_{-,m} \right].
\label{BCH-evolution}
\end{equation}
The coefficients above have the form
\begin{align}
\chi_{+,m} (t_1 , t_2 ) & =  \frac{\phi^{(2)}_{m} (t_1 , t_2 ) \sinh(\Delta(t_1 , t_2 ))}{\Delta(t_1 , t_2 ) \Lambda (t_1 , t_2 ) } ,
\label{chi-plus}
\\
\chi_{0,m} (t_1 , t_2 ) & =  - 2 \ln 
\left[
\Lambda (t_1 , t_2 )
\right] ,
\label{chi-zero}
\\
\chi_{-,m} (t_1 , t_2 ) & =  \frac{\phi^{(1)}_{m} (t_1 , t_2 ) \sinh(\Delta(t_1 , t_2 ))}{\Delta(t_1 , t_2 ) \Lambda (t_1 , t_2 ) } ,
\label{chi-minus}
\end{align}
where we have used the following shorthand notations
\begin{align}
\Lambda (t_1 , t_2 ) & =  \cosh(\Delta(t_1 , t_2 )) - \frac{\omega_{m}(t_1 , t_2 )}{\Delta(t_1 , t_2 )} \sinh(\Delta(t_1 , t_2 )),
\label{chi-zero-1}
\\
\Delta(t_1 , t_2 ) & =  \omega^{2}_{m} (t_1 , t_2 ) - 
\phi^{(1)}_{m} (t_1 , t_2 ) 
\phi^{(2)}_{m} (t_1 , t_2 ) .
\label{Delta}
\end{align}
It is important to observe that the coefficients $\chi_{+,m} (t_1 , t_2 )$, $\chi_{0,m} (t_1 , t_2 )$ and $\chi_{-,m} (t_1 , t_2 )$ have this form only if the integral from $\normord{\Upsilon (t_2,t_1)}$ is definite. In this case, the coefficients we start with, namely $\omega_{m} (t_1 , t_2 )$, $\phi^{(1)}_{m} (t_1 , t_2 )$ and $\phi^{(2)}_{m} (t_1 , t_2 )$, are the definite integrals of the corresponding functions $\Omega_{m} (t)$, $\Phi^{(1)}_{m} (t)$ and $\Phi^{(2)}_{m} (t)$, respectively. If one chooses to work with indefinite integrals instead, then the coefficient $\chi_{0,m} (t)$ does not have a general closed form. 

The density matrix can be constructed by plugging the BCH relation (\refeq{BCH-evolution}) into the equation (\refeq{unitary-map}) and using the result in 
\begin{equation}
\rho_{\Psi'} (t_1 , t_2) = | \Psi' (t_2) \rangle \langle \Psi' (t_2) |
= \normord{\Upsilon (t_2,t_1)}| \Psi (t_1) \rangle 
\langle \Psi (t_1) | \normord{\Upsilon (t_2,t_1)}^{\dagger}.
\label{density-matrix-calc}
\end{equation} 
The LREE in an arbitrary state $\rho_{\Psi'} (t_1 , t_2)$ is the von Neumann entropy of the reduced density matrix obtained by tracing out the degrees of freedom of either $\alpha$ or $\beta$ modes, respectively. The result depends on the initial state $| \Psi (t_1) \rangle$ one starts with, but since an arbitrary state contains an arbitrary number of oscillators, the entanglement entropy is, in general, indefinitely large. 

Let us particularize the equation (\refeq{density-matrix-calc}) to the ground state. Since the different modes are independent, we will focus on a single mode in one direction. Then the evolved background state has the following form in the BF representation
\begin{equation}
| \Psi_{0,m} (t_1, t_2) \rangle = 
\exp \left[ 
\frac{1}{2} \chi_{0,m} (t_1 , t_2 ) 
\right]
\sum_{n_{m}=0}^{\infty}
\left( \chi_{+,m} (t_1 , t_2 ) \right)^{n_{m}} 
|n_{m} \rangle_{\alpha} |n_{m} \rangle_{\beta} .
\label{evolved-background}
\end{equation}
Recall that the von Neumann entropy in the state $\Psi$ is defined in terms of the density matrix $ \rho_{\Psi}$ associated to the state by the following relation
\begin{equation}
S[\rho_{\Psi}] = - \mbox{Tr} \left[ \rho_{\Psi} \ln \rho_{\Psi} \right].
\label{von-Neumann-entropy}
\end{equation}
Now if the system is bipartite, the reduced density matrix of one subsystem is defined by tracing over the degrees of freedom of the other subspace. Then the entanglement entropy of the system is defined as the von Neumann entropy calculated for the reduced density matrix \cite{Nielsen:2011}. If we apply this  
definition to the $\sigma$-model, it is easy to see that by tracing over the $\alpha$ degrees of freedom and by applying the formula (\refeq{von-Neumann-entropy}) we obtain the following entanglement entropy 
\begin{equation}
S_{\beta} [\Psi_{0,m} (t_1 , t_2)] = - \sum_{n_m = 0}^{\infty}
e^{\frac{1}{2} \mathrm{Re} (\chi_{0,m} (t_1 , t_2 ))}
|\chi_{+,m} (t_1 , t_2 ) |^{n_m}
\ln 
\left[
e^{\frac{1}{2} \mathrm{Re} (\chi_{0,m} (t_1 , t_2 ))}
|\chi_{+,m} (t_1 , t_2 ) |^{n_m}
\right].
\label{relative-entropy-evolved-background}
\end{equation} 
On symmetry grounds as well as by direct calculations, we can infer that 
\begin{equation}
S_{\beta} [\Psi_{0,m} (t_1 , t_2)] = S_{\alpha} [\Psi_{0,m} (t_1 , t_2)].
\label{S-S}
\end{equation}
This shows that the evolved ground state of a single mode in the BS representation is a pure state. It is straightforward to write down the full entanglement entropy when all bi-oscillators are taken into account. Since the modes are independent, this is just the double product over all indices $i$ and $m$, respectively, of the entanglement entropy calculated above. 

\subsection{Asymptotic Entanglement Entropy of a Single Mode}

In this subsection we want to calculate explicitly the asymptotic form of the entanglement entropy given by the equation (\refeq{relative-entropy-evolved-background}) at large values of times $t_1 \rightarrow \infty$ and $t_2 \rightarrow \infty$, respectively, while keeping the interval $\Delta t = t_2 - t_1$ constant. To that end, we employ the asymptotic representation of the Bessel and Hankel functions \cite{Olver:2010} summarized for convenience in Appendix B. Also, we will consider the phenomenological approximation $H\Delta t \ll 1$ which is valid even for larger values of $\Delta t$ as $H \sim 10^{-18}s^{-1}$.

From the definition (\refeq{Omega}) and the asymptotic relations (\refeq{Y0-Y1}), after some lengthy but straightforward computations, we obtain the following form of the integrals of the functions $\Omega_{m} (z_m)$, 
$\Phi^{(1)}_{m} (z_m)$ and $\Phi^{(2)}_{m} (z_m)$ between $t_1$ and $t_2$
\begin{align}
\omega(t_1 ,t_2) & \sim \frac{5m}{2H} H \Delta t
+
\frac{3}{2} \left[
\cos\left( \frac{2m}{H}e^{Ht_2} \right)
-
\cos\left( \frac{2m}{H}e^{Ht_1} \right)
\right],
\label{asympt-omega}
\\
\phi^{(1)}_{m} (t_1,t_2) & \sim \frac{5H}{2} H \Delta t
-
\frac{3i}{2}
\left[
E_{1} \left( \frac{2mi}{H}e^{Ht_2} \right)
-
E_{1} \left( \frac{2mi}{H}e^{Ht_1} \right)
\right],
\label{asympt-fi1}
\\
\phi^{(2)}_{m} (t_1,t_2) & \sim \frac{5H}{2} H \Delta t
+
\frac{3i}{2}
\left[
E_{1} \left( \frac{-2mi}{H}e^{Ht_2} \right)
-
E_{1} \left( \frac{-2mi}{H}e^{Ht_1} \right)
\right],
\label{asympt-fi2}
\end{align}
where $E_1(z)$ is the exponential integral. Since $H\Delta t$ and $H^2 \Delta t^2$ can be neglected when compared with $\exp(Ht_1)$ in the limits under consideration, one can simplify further the above relations. By substituting these results into the equations (\refeq{chi-plus}) and (\refeq{chi-zero}) we obtain the asymptotic expressions of the  
the parameters $\chi_{+,m} (t_1 , t_2 )$ and $\chi_{0,m} (t_1 , t_2 )$, respectively,
\begin{align}
\chi_{+,m} (t_1 , t_2 ) & \sim  
\frac{2H}{5m^2}\frac{e^{-Ht_1}}{\Delta t} 
\tanh 
\left( 
\frac{25}{4}m^2 e^{2Ht_1} \Delta t^2
\right),
\label{asympt-chi-plus}\\
\chi_{0,m} (t_1 , t_2 ) & \sim
-2 \ln \left[ 
\cosh \left(
\frac{25}{4}m^2 e^{2Ht_1} \Delta t^2
\right)
\right] .
\label{asympt-chi-zero}
\end{align}
By using the above equations into the formula (\refeq{relative-entropy-evolved-background}), we obtain the asymptotic LREE in the evolved background state 
\begin{align}
S_{\beta} [\Psi_{0,m} (t_1 , t_2)] & \sim  - \sum_{n_m = 0}^{\infty}
\left[
\cosh \left(
\frac{25}{4}m^2 e^{2Ht_1} \Delta t^2
\right)
\right]^{-1}
\left|
\frac{2H}{5m^2}\frac{e^{-Ht_1}}{\Delta t} 
\tanh 
\left( 
\frac{25}{4}m^2 e^{2Ht_1} \Delta t^2
\right)
\right|^{n_m}
\nonumber
\\
& \times
\ln
\left[
\left[
\cosh \left(
\frac{25}{4}m^2 e^{2Ht_1} \Delta t^2
\right)
\right]^{-1}
\left|
\frac{2H}{5m^2}\frac{e^{-Ht_1}}{\Delta t} 
\tanh 
\left( 
\frac{25}{4}m^2 e^{2Ht_1} \Delta t^2
\right)
\right|^{n_m}
\right].
\label{relative-entropy-background-asympt}
\end{align}
Note that the asymptotic expansion at $t \rightarrow 0$ is inaccessible to every oscillating mode. Indeed, from the definition of the parameter $z_m$ given by the equation (\refeq{notation-zm}) we see that in this limit the Bessel functions are calculated at $m/H$.

\section{Discussions}

We have obtained the formula of the entanglement entropy between the left and right modes in the ground state of the $\sigma$-model with the de Sitter target space and have explicitly determined its asymptotic representation at large values of time. Our findings have been obtained in the cosmological gauge that is the time-dependent gauge compatible with the canonical quantization relations. In this sense, the cosmological gauge is natural to the theory.
Also, we have discussed the classical dynamics of the $\sigma$-model in this gauge and have found a new general solution to the equations of motion. This solution is a superposition of Hankel functions that reflect the symmetry of the embedded base space. More important, the oscillators are time-independent. We have studied the action of the constraints on the space of states in the Gupta-Bleuler quantization and from that we have argued that the Hilbert space can be expressed as a (time-parametrized) one-parameter family of Hilbert spaces that is also a direct product of smooth time-dependent functions with the time-independent Fock space of all excitations. Then, we have shown that the Hilbert space of the system is an infinite-dimensional direct product of irreducible representations of the $SU(1,1)$ group in the Jordan-Schwinger representation. The physical subspace has been determined in this representation. An important map between the states of the $\sigma$-model at different values of time is provided by the Hamiltonian (the partial evolution map). We have clarified the relationship between this map and the time-evolution of the total system that includes the de Sitter background and have shown that its main effect on states is to alter the entanglement between the left and right oscillating modes during a finite interval of time. This entanglement is measured by the entanglement entropy of the evolved ground state which we have determined by using the Jordan-Schwinger representation. Also, we have derived its asymptotic form at large values of time for an arbitrary mode.

Compared with the literature, the LREE defined and calculated in the present paper is a physical property of the free $\sigma$-model fields, rather than of the boundary states of the string theory as mentioned in the introduction. Another interesting observation is that 
similar operators to the partial evolution map have been constructed in the context of entanglement renormalization within the holographic principle in \cite{Haegeman:2011uy,Molina-Vilaplana:2015mja,Molina-Vilaplana:2015rra,Caputa:2017urj}\footnote{I acknowledge the anonymous referee for pointing out to me the similarity between the two operators.}. On general grounds, the existence of the Hamiltonian-like operators in these two different contexts is not a surprise. Indeed, the local (in the target space) evolution of the $\sigma$-model is similar to the renormalization flow along the time-like Killing line. Another way to see that is by diagonalising the $\sigma$-model Hamiltonian (\refeq{quantum-Hamiltonian}) and constructing its double field theory. The system obtained in this way has properties of a non-equilibrium field theory \cite{Vancea:2017pom}. This is an interesting fact, as it shows that the connection between the entanglement entropy of the $\sigma$-model in the de Sitter space and the thermal field theory could be defined in the non-equilibrium field theory. 

Let us briefly discuss how the LREE obtain in this paper could help with the general problems given in the introduction to justify the study of the $\sigma$-model in the de Sitter space-time. As we have seen, the LREE is an important physical quantity that contains information not only about the quantum $\sigma$-model but also on the action of the space-time metric on the physical quanta. Unlike the flat space-time, even if the $\sigma$-model degrees of freedom do not interact among themselves (free fields) they still interact with the metric components.  
While these is true for all quantum objects constructed out of fields, the  LREE is particularly interesting since one could in principle compute the entropy spectrum which characterizes the topological nature of the quantum $\sigma$-model, e. g. the gaps in the entanglement and the topological phases. Since the components of the metric determine these phases, the entropy spectrum contains information about how the background determines the quantum topology of the system. The same idea can be applied to understand the microscopic properties of more general time-dependent backgrounds in which the entanglement entropy is computable, e. g. \cite{Madhu:2009jh}. Therefore, the LREE can be used to understand the microscopic structure of these backgrounds through their interactions with the $\sigma$-model which can be viewed as quantum probe. From this point of view, the $\sigma$-model plays a singular role in investigating the space-time microscopic properties in contrast with other field theories that have a different relationship with the space-time events. The present analysis could be generalized to the non-critical string compactified to a space-time model with a cosmological constant. In that case, the extra fields (like the dilaton) necessary to guarantee the Weyl symmetry and the compactified modes will also be entangled with the left and right moving modes and will determine the topological properties of the non-critical string
through the LREE. While technically more difficult, the steps performed here should be derivable in that case, too. We hope to report on these topics elsewhere.

An interesting avenue to be explored is the interpretation of the LREE in terms of space-time quantities, mainly whether there is any relationship between the LREE and the entanglement entropy of a given region of space-time. Such of question should be posed in a more concrete formulation like KKLT or MERA in which the de Sitter background can be related in principle with the microscopic string.
In the present case, the answer seems to be in negative since the quantum $\sigma$-model does not determine the de Sitter space-time in the same way as the string does with the Minkowski space-time due to the fact that the de Sitter space is not a string background. 

Returning to the $\sigma$-model in the de Sitter space-time, there are several issues to be solved from the physical point of view. The first one is to determined the correlators among the oscillating modes of the $\sigma$-model. This problem is complicated by the mutual interaction of the oscillators and of the background. The second problem is to describe the thermal effects related to the expansion of the background which has been partially solved in \cite{Vancea:2017pom}. A third issue is related to the fact that the results obtained here are gauge-dependent and local in the de Sitter space, in the sense that they are derived in a neighbourhood of the embedded cylinder in a patch of the full space. However, they are relevant because of the conceptual importance of the cosmological gauge stressed above. Also, since there is no known method to study the quantum systems covariantly and globally in the de Sitter background at present, the gauge fixed theory provides the only information about the $\sigma$-model. Nevertheless, it would be interesting to see how the results obtained here could help to formulate a more covariant theory of the $\sigma$-model in the de Sitter space as well as to understand the microscopic structure of this phenomenological background and of the quantum dynamics of the (non-critical) string theory on it.

\section*{Acknowledgements}
I would like to thank to A. Sen for very valuable discussions. I also enjoyed conversation with I. Sachs. It is a pleasure to acknowledge N. Berkovits for hospitality at ICTP-SAIFR where a significant part of this work was completed. Also, I acknowledge the anonymous referee for helping me improve greatly the text.

\section*{Appendix A}

In this appendix, we discuss the canonical quantization of the general solution to the equations of motion of the $\sigma$-model in an arbitrary time-dependent gauge $\omega(t)$. This solution was obtained previously in \cite{Li:2007gf}. 

For a general diagonal gauge given by the equations (\refeq{static-gauge}) and (\refeq{metric-gauge-ws}), the equations of motion take the form
\begin{equation}
\partial_t \left[ e^{2Ht} \omega(t) \partial_t x^{i}(t, \sigma)\right]
- e^{2Ht} \partial_{\sigma} \left[ \omega^{-1}(t) \partial_{\sigma} x^{i}(t, \sigma) \right] =0.
\label{eq-mot-arbitrary}
\end{equation}
This equation can be solved to obtain the following general solution
\begin{equation}
x^{i} (t,\sigma)  =  x^{i}_{0} + p^{i}_{0} \int^{t} dy f^{2}(y)
+ \sum_{m > 0} \frac{f(t)}{\sqrt{2 |g_{m}(t)|}} 
\left[ 
\alpha^{i}_{m} (t) e^{im \sigma} + \beta^{i}_{m} (t) e^{-im \sigma} 
\right] F_{m} (t).
\label{gen-sol-arbitrary}
\end{equation}
Thus, in the arbitrary time-dependent gauge, the oscillating excitations 
$\alpha^{i}_{m} (t)$ and  $\beta^{i}_{m} (t)$ depend on time. The functions $F_{m} (t)$ are given by the following relation
\begin{equation}
F_{m} (t) = \exp \left[ -i \int^{t} dy  g_{m}(y)\right],
\label{Fm-function}
\end{equation}
where we have used the notations
\begin{equation}
f(t) = \omega^{-\frac{1}{2}}(t) e^{-Ht},
\hspace{0.5cm}
g_{m} (t) = \mbox{sgn}(m)
\left[ 
\left( \frac{m}{\omega(t)}\right)^{2}
- f(t) 
\frac{d^2}{dt^2}\left( \frac{1}{f(t)} \right)
\right]^{\frac{1}{2}}.
\label{f-gamma}
\end{equation}
The canonical quantization requires that the fields $x^{i} (t,\sigma)$ and their canonically conjugate momenta $\pi^i(t, \sigma)$ satisfy the commutation relations (\refeq{ETC}) while the modes $\alpha^{i}_{m} (t)$ and  $\beta^{i}_{m} (t)$ satisfy the oscillator commutators
\begin{equation}
[ \alpha^{i}_{m} (t), \alpha^{j}_{n} (t) ] = 
[ \beta^{i}_{m} (t), \beta^{j}_{n} (t) ] = \frac{m}{|m|}\delta^{ij} \delta_{mn}
.
\label{comm-rel-li}
\end{equation}
However, the operators $x^{i} (t,\sigma)$ and $\pi^i(t, \sigma)$ depend on time while the right hand side of the equation (\refeq{ETC}) is time-independent. A short computation in which the equations (\refeq{ETC}) and (\refeq{gen-sol-arbitrary}) are used show that the general solution given by the equation (\refeq{gen-sol-arbitrary}) satisfy the canonical commutation relations only if the following equation is satisfied 
\begin{equation}
g^{-1}_{m} (t) \left[ \frac{d}{d t} \left( \ln f(t) \right) + i g_m (t) \right] = C,
\label{time-dep-ETC}
\end{equation}
for all $m > 0$. The solution to the above equation (\refeq{time-dep-ETC}) is $\omega (t) = \omega_0 \exp(-2Ht)$ which is the gauge parameter of the cosmological gauge. The value of $\omega_0$ is fixed to one by the same equation (\refeq{ETC}) and the condition that the delta-function has the standard normalization. Thus, we can conclude that the cosmological gauge defined by the equation (\refeq{metric-cosmological-gauge}) is a consequence of the quantization of the classical $\sigma$-model in a time-dependent arbitrary gauge. Moreover, it is the time-dependent gauge in which the canonically quantization relations are consistent with the oscillator interpretation of string excitations.

\section*{Appendix B}

The asymptotic form of the Bessel and Hankel functions is presented in \cite{Olver:2010}. We reproduce here the formulas relevant for the computations of the asymptotic form of the entanglement entropy at $z \rightarrow \infty$ and fixed value of $\nu = 0$.

The asymptotic representation of the Bessel functions of the second kind truncated to the coefficients of $z^{-1}$ is given by the following equations
\begin{equation}
Y_{0} (z)  \sim  \sqrt{\frac{2}{\pi z}} 
\sin (z - \frac{\pi}{4}),  
\hspace{0.5cm}
Y'_{0} (z)  \sim  \sqrt{\frac{2}{\pi z}} 
\cos (z - \frac{\pi}{4}).
\label{Y0-Y1}
\end{equation}
Also, we use the following asymptotic representation of the Hankel functions 
\begin{align}
H^{(1)}_{0} (z) & \sim  \sqrt{\frac{2}{\pi z}} e^{i (z - \frac{\pi}{4})},
\label{H1-asympt}
\\
H^{(2)}_{0} (z)  & \sim \sqrt{\frac{2}{\pi z}}  e^{- i (z - \frac{\pi}{4})},
\label{H2-asympt}
\\
H^{(1)'}_{0}  (z) & \sim  i \sqrt{\frac{2}{\pi z}} e^{i (z - \frac{\pi}{4})},
\label{H1-deriv-asympt}
\\
H^{(2)'}_{0}  (z) & \sim  -i \sqrt{\frac{2}{\pi z}} e^{- i (z - \frac{\pi}{4})}.
\label{H2-deriv-asympt}
\end{align}


\begin{thebibliography}{Proper}


\bibitem{Muller:1995mz} 
  R.~Muller and C.~O.~Lousto,
  Phys.\ Rev.\ D {\bf 52}, 4512 (1995).
  \doi{10.1103/PhysRevD.52.4512}

\bibitem{Kabat:2002hj} 
  D.~N.~Kabat and G.~Lifschytz,
  JHEP {\bf 0204}, 019 (2002).
  \doi{10.1088/1126-6708/2002/04/019}

\bibitem{Maldacena:2012xp} 
  J.~Maldacena and G.~L.~Pimentel,
  JHEP {\bf 1302}, 038 (2013).
  \doi{10.1007/JHEP02(2013)038}

\bibitem{Nomura:2013lia} 
  Y.~Nomura and S.~J.~Weinberg,
  Phys.\ Rev.\ D {\bf 90}, no. 10, 104003 (2014).
  \doi{10.1103/PhysRevD.90.104003}

\bibitem{Kanno:2014lma} 
  S.~Kanno, J.~Murugan, J.~P.~Shock and J.~Soda,
  JHEP {\bf 1407}, 072 (2014).
  \doi{10.1007/JHEP07(2014)072}

\bibitem{Iizuka:2014rua}
  N.~Iizuka, T.~Noumi and N.~Ogawa,
  Nucl.\ Phys.\ B {\bf 910} (2016) 23.
  \doi{10.1016/j.nuclphysb.2016.06.024}

\bibitem{Pavao:2016rhe} 
  R.~Pavão, R.~Faleiro, A.~H.~Blin and B.~Hiller,
  arXiv:1607.02115 [gr-qc].


\bibitem{Schmidt:1998ys} 
  B.~P.~Schmidt {\it et al.} [Supernova Search Team Collaboration],
  Astrophys.\ J.\  {\bf 507}, 46 (1998).
  \doi{10.1086/306308}

\bibitem{Riess:1998cb} 
  A.~G.~Riess {\it et al.} [Supernova Search Team Collaboration],
  Astron.\ J.\  {\bf 116}, 1009 (1998).
  \doi{10.1086/300499}

\bibitem{Perlmutter:1998np} 
  S.~Perlmutter {\it et al.} [Supernova Cosmology Project Collaboration],
  Astrophys.\ J.\  {\bf 517}, 565 (1999).
  \doi{10.1086/307221}


\bibitem{Maloney:2002rr} 
  A.~Maloney, E.~Silverstein and A.~Strominger,
  hep-th/0205316.

\bibitem{Burgess:2003ic} 
  C.~P.~Burgess, R.~Kallosh and F.~Quevedo,
  JHEP {\bf 0310}, 056 (2003).
  \doi{10.1088/1126-6708/2003/10/056}

\bibitem{Kachru:2003aw} 
  S.~Kachru, R.~Kallosh, A.~D.~Linde and S.~P.~Trivedi,
  Phys.\ Rev.\ D {\bf 68}, 046005 (2003).
  \doi{10.1103/PhysRevD.68.046005}


\bibitem{Gubser:2003vk} 
  S.~S.~Gubser,
  Phys.\ Rev.\ D {\bf 69}, 123507 (2004).
  \doi{10.1103/PhysRevD.69.123507}


\bibitem{deRham:2015ijs} 
  C.~de Rham, A.~J.~Tolley and S.~Y.~Zhou,
  Phys.\ Lett.\ B {\bf 760}, 579 (2016).
  \doi{10.1016/j.physletb.2016.07.035}

\bibitem{deRham:2016plk} 
  C.~de Rham, A.~J.~Tolley and S.~Y.~Zhou,
  JHEP {\bf 1604}, 188 (2016),
  \doi{10.1007/JHEP04(2016)188}



\bibitem{deVega:1987veo} 
  H.~J.~de Vega and N.~G.~Sanchez,
  Phys.\ Lett.\ B {\bf 197}, 320 (1987).
  \doi{10.1016/0370-2693(87)90392-3}

\bibitem{deVega:1994yz} 
  H.~J.~de Vega, A.~L.~Larsen and N.~G.~Sanchez,
  Phys.\ Rev.\ D {\bf 51}, 6917 (1995).
  \doi{10.1103/PhysRevD.51.6917}

\bibitem{RamonMedrano:1999gm} 
  M.~Ramon Medrano and N.~G.~Sanchez,
  Phys.\ Rev.\ D {\bf 60}, 125014 (1999).
  \doi{10.1103/PhysRevD.60.125014}

\bibitem{Bouchareb:2005ck} 
  A.~Bouchareb, M.~Ramon Medrano and N.~G.~Sanchez,
  Int.\ J.\ Mod.\ Phys.\ D {\bf 16}, 1053 (2007).
  \doi{10.1142/S0218271807010596}
 
\bibitem{Li:2007gf} 
  M.~Li, W.~Song and Y.~Song,
  JHEP {\bf 0704}, 042 (2007)
  \doi{10.1088/1126-6708/2007/04/042}
  [hep-th/0701258].


\bibitem{Viswanathan:1996yg} 
  K.~S.~Viswanathan and R.~Parthasarathy,
  Phys.\ Rev.\ D {\bf 55}, 3800 (1997)
  \doi{10.1103/PhysRevD.55.3800}
  [hep-th/9605007].

\bibitem{Bozhilov:2001kw} 
  P.~Bozhilov,
  Phys.\ Rev.\ D {\bf 65}, 026004 (2002)
  \doi{10.1103/PhysRevD.65.026004}
  [hep-th/0103154].
  

\bibitem{Callan:1985ia} 
  C.~G.~Callan, Jr., E.~J.~Martinec, M.~J.~Perry and D.~Friedan,
  Nucl.\ Phys.\ B {\bf 262}, 593 (1985).
  \doi{10.1016/0550-3213(85)90506-1}

\bibitem{Vidal:2008zz} 
  G.~Vidal,
  Phys.\ Rev.\ Lett.\  {\bf 101}, 110501 (2008).
  \doi{10.1103/PhysRevLett.101.110501}

\bibitem{Haegeman:2011uy} 
  J.~Haegeman, T.~J.~Osborne, H.~Verschelde and F.~Verstraete,
  Phys.\ Rev.\ Lett.\  {\bf 110}, no. 10, 100402 (2013)
  \doi{10.1103/PhysRevLett.110.100402}
  [arXiv:1102.5524 [hep-th]].

\bibitem{Molina-Vilaplana:2015mja} 
  J.~Molina-Vilaplana,
  JHEP {\bf 1509}, 002 (2015)
  \doi{10.1007/JHEP09(2015)002}
  [arXiv:1503.07699 [hep-th]].

\bibitem{Das:2015oha} 
  D.~Das and S.~Datta,
  Phys.\ Rev.\ Lett.\  {\bf 115}, no. 13, 131602 (2015)
  \doi{10.1103/PhysRevLett.115.131602}
  [arXiv:1504.02475 [hep-th]].
 
\bibitem{Molina-Vilaplana:2015rra} 
  J.~Molina-Vilaplana,
  Phys.\ Lett.\ B {\bf 755}, 421 (2016)
  \doi{10.1016/j.physletb.2016.02.050}
  [arXiv:1510.09020 [hep-th]].

\bibitem{Caputa:2017urj} 
  P.~Caputa, N.~Kundu, M.~Miyaji, T.~Takayanagi and K.~Watanabe,
  Phys.\ Rev.\ Lett.\  {\bf 119}, no. 7, 071602 (2017)
  \doi{10.1103/PhysRevLett.119.071602}
  [arXiv:1703.00456 [hep-th]].
 


\bibitem{PandoZayas:2014wsa} 
  L.~A.~Pando Zayas and N.~Quiroz,
  JHEP {\bf 1501}, 110 (2015)
  \doi{10.1007/JHEP01(2015)110}
  [arXiv:1407.7057 [hep-th]].

\bibitem{Zayas:2016drv} 
  L.~A.~Pando Zayas and N.~Quiroz,
  JHEP {\bf 1611}, 023 (2016)
  \doi{10.1007/JHEP11(2016)023}
  [arXiv:1605.08666 [hep-th]].

\bibitem{Schnitzer:2015gpa} 
  H.~J.~Schnitzer,
  arXiv:1505.07070 [hep-th].


\bibitem{Madhu:2009jh} 
  K.~Madhu and K.~Narayan,
  Phys.\ Rev.\ D {\bf 79}, 126009 (2009)
 \doi{10.1103/PhysRevD.79.126009}
  [arXiv:0904.4532 [hep-th]].


\bibitem{Bargmann:1946me} 
  V.~Bargmann,
  Annals Math.\  {\bf 48}, 568 (1947).
  \doi{10.2307/1969129}


\bibitem{Olver:2010}
F.~ W.~ J.~ Olver,
D.~ W.~ Lozier,
R.~ F.~ Boisvert,
C.~ W.~ Clark,
NIST Handbook of Mathematical Functions, Cambridge University Press, 2010.

\bibitem{Jordan:1935} 
  P.~Jordan 
  Z.\ Phys.\  {\bf 94}, 531 (1935).
  
\bibitem{Schwinger:1965}
J. ~Schwinger, in: Quantum Theory of Angular Momentum, 
eds. L. C. Biedenharn and H. van Dam, Academic Press, New York (1965), p. 229.


\bibitem{Perelomov:1986tf} 
  A.~M.~Perelomov,
  Generalized coherent states and their applications,
  Berlin, Germany: Springer (1986).
  

\bibitem{Shaterzadeh:2008}
Z.~ Shaterzadeh-Yazdi, P.~ S.~ Turner and B.~ C.~ Sanders,
J.\ Phys. \ A: Math. \ Theor. {\bf 41} (2008) 055309,
[arXiv:0710.3205 [quant-ph]].  


\bibitem{Nielsen:2011}
M. ~ A. ~ Nielsen and I. ~ L. ~ Chuang,
Quantum Computation and Quantum Information,
Cambridge University Press, 2000.




\bibitem{Wilcox:1967zz} 
  R.~M.~Wilcox,
  J.\ Math.\ Phys.\  {\bf 8}, 962 (1967).
  \doi{10.1063/1.1705306}

\bibitem{Gilmore:1974}
	R.~ Gilmore,
	J.\ Math.\ Phys. {\bf 15}, 2090 (1974).

\bibitem{Truax:1985vk} 
  D.~R.~Truax,
  Phys.\ Rev.\ D {\bf 31}, 1988 (1985).
  \doi{10.1103/PhysRevD.31.1988}
  
\bibitem{Vancea:2017pom} 
  I.~V.~Vancea,
  Adv.\ High Energy Phys.\  {\bf 2017}, 3706870 (2017)
  \doi{10.1155/2017/3706870}
  [arXiv:1701.05582 [hep-th]].

\end{thebibliography}
\end{document}